\documentclass[12pt,preprint]{aastex}

\shorttitle{Arecibo  430 MHz Pulsar Polarimetry}
\shortauthors{Weisberg et al}

\begin{document}

\title{Arecibo 430 MHz Pulsar Polarimetry: Faraday Rotation Measures and 
Morphological Classifications}

\author{J.M.~Weisberg\altaffilmark{1}, J.M.~Cordes\altaffilmark{2},
B. Kuan\altaffilmark{1}, K.E.~Devine\altaffilmark{1}, 
J.T.~Green\altaffilmark{1}, and  D.C.~Backer\altaffilmark{3} }
\altaffiltext{1} {Department of Physics and Astronomy, Carleton College,
Northfield, MN 55057, \it{jweisber@carleton.edu}}
\altaffiltext{2} {Astronomy Department and NAIC, Cornell University, Ithaca,
NY 14853, \it{cordes@spacenet.tn.cornell.edu}}
\altaffiltext{3} {Astronomy Department and Radio Astronomy Laboratory, 
University of California, Berkeley, CA 94720-3411, \it{dbacker@astro.berkeley.edu}}


\email{jweisber@carleton.edu}


%

\begin{abstract}
We have measured Faraday Rotation Measures $(RM)$ at Arecibo Observatory for 
36 pulsars, 17 of them new. 
We combine these and earlier measurements to study the galactic magnetic field 
and its possible temporal variations.  Many $RM$ values have changed significantly
on several--year timescales, but these variations probably do not reflect 
interstellar magnetic field changes.  By studying the distribution of pulsar $RM$s 
near the plane in conjunction with the new NE2001 electron density model,
we note the following structures in the first galactic longitude quadrant:  
(1) The local field reversal can be traced as a null in 
$RM$ in a 0.5--kpc wide strip interior to the Solar Circle, extending $\sim7$ kpc 
around the Galaxy. (2) Steadily increasing $RM$s in a 1--kpc wide strip interior to the 
local field reversal, and also in the wedge bounded by $42<l<52\degr$, indicate that 
the large--scale field is approximately steady from the local reversal in to the
Sagittarius arm.  (3) The $RM$s in the 1--kpc wide strip interior to the Sagittarius arm
indicate another field reversal in this strip. (4) The $RM$s in a final 1--kpc wide 
interior strip, 
straddling the Scutum arm, also support a second field reversal interior to the Sun,
between the Sagittarius and Scutum arms.  (5) Exterior to the nearby reversal, $RM$s 
from
$60<l<78\degr$ show evidence for {\em{two}} reversals, on the near and far side of
the Perseus arm. (6) In general, the maxima in the 
large--scale fields tend to lie along the spiral arms, while the field minima tend
to be found between them. 

We have also determined polarized profiles of 48 pulsars at 430 MHz.
We  present morphological pulse profile classifications \citep{R83} of the 
pulsars, based on our new measurements and previously published data. 
\end{abstract}


\section{INTRODUCTION}

We have determined full--polarization properties of 48 pulsars at 430 MHz from 
Arecibo
Observatory.  This paper presents    the principal 
results of these measurements--Faraday rotation and interstellar magnetic 
field determinations, and polarized profiles and pulsar morphological 
classifications.  The plan of the paper is as follows:  In the remainder of
the Introduction, we summarize the background on these two primary results.
The second section of the paper describes the observations and our analyses
of them.  The third section discusses our Faraday rotation and galactic magnetic
field results in detail, while the fourth describes the polarized profiles
and morphological classifications.  Conclusions are given in the fifth section.

\subsection{Faraday Rotation and Interstellar Magnetic Field Measurements}

Pulsars are ideal probes of the galactic magnetic field.  Faraday Rotation 
measurements of pulsars' linearly polarized emissions yield the Rotation 
Measure, 
$RM$, which is a path integral along the line of sight involving the magnetic
field $B$ and electron density $n_e$:

\begin{equation}
RM=0.81\int\limits_{PSR}^\earth  {n_e {\bf B}\cdot d{\bf s} }   ;
\end{equation}
while pulse timing measurements determine the Dispersion Measure, $DM$,
 which
is another path integral along the line of sight involving electron 
density $n_e$ alone:

\begin{equation}
DM=\int\limits_{PSR}^\earth  {n_e ds }.   
\end{equation}

Consequently the mean value of the component of the magnetic field along the
line of sight, $<B_{||}>$, is given by:

\begin{equation}
\left\langle {B_{||}} \right\rangle =\int\limits_{PSR}^\earth  
{n_e {\bf B}\cdot d{\bf s}\; }/\int\limits_{PSR}^\earth  {n_eds\;}
=RM / (0.81DM).
\end{equation}

The galactic magnetic field structure is particularly amenable to study
with pulsars because of this relationship, which can be evaluated with pulsars
at various distances along a particular line of sight.  \citet{ID98} and
\citet{H99}  provide recent comprehensive analyses of the galactic magnetic
field.

\subsection{Pulsar Morphological Classifications}

\citet{R83} proposed a morphological pulsar classification scheme that was
based on examination of multifrequency, polarized pulse profiles.  Rankin
found that there are two principal classes of emission beams --  core and 
conal,
whose properties are distinguishable through careful analysis of 
polarized properties as a function of frequency.   She and others have 
applied her classification criteria to large number of pulsars in subsequent 
papers  (Rankin 1986, 1990, 1993; Rankin, Stinebring, \& Weisberg 1989; 
Weisberg et al 1999). In the current paper, we measure a large number of 
polarized profiles
at 430 MHz, and use them and earlier multifrequency measurements to make
new morphological classifications and to improve old ones.

\section{OBSERVATIONS AND ANALYSIS}

The data were gathered with the dual--circularly polarized 430 MHz linefeed 
at Arecibo Observatory. A 5 MHz band centered at 430 MHz was fed into the 
Arecibo correlator, and 32 lags of auto-- and 
cross--correlation functions were formed every  $400~\mu s$.
Using an ephemeris for the apparent pulsar period generated by program
{\sc TEMPO} \citep{TW89} to calculate the pulsar rotational phase at the epoch
of each such sample, we selected one of 1024 pulse longitude bins to accumulate 
the resulting four correlation functions. This procedure continued
for 120 seconds of observing. Off--line, the four correlation functions  at 
each longitude bin were three--level corrected and Fourier transformed 
to form   pulse profiles in the four Stokes parameters.
This procedure was essentially identical to that used in the 21 cm 
observations of \citet{W99}, except that  the 430 MHz linefeed limited 
the bandwidth to a narrower range.  At this point in the processing, the 
data consisted of a cube of 1024 longitude bins x 4 Stokes parameters x 32 
frequency channels of 156.25 kHz bandwidth, representing a two minute 
integration or ``scan.''

\subsection{Faraday Rotation Measurements}

Observationally speaking, the Faraday Rotation Measure  $RM$ is conventionally 
expressed as a rotation of the linearly polarized position angle $\psi$ 
as a function
of {\em{wavelength}} $\lambda$:
\begin{equation}
\psi-\psi_0 = RM \;\lambda^2;
\end{equation}
where $RM$ has units of rad m$^{-2}$.  The rate of rotation with wavelength
is then
\begin{equation}
{d\psi \over d\lambda}  = 2\; RM\; \lambda
\end{equation}
  It is more
convenient for us initially to measure it as a position angle rotation rate 
with {\em{frequency}}:
\begin{equation}
{d\psi \over d\nu} =-{2\;RM\;c^2\;\over\;\nu^3}
\end{equation}


At our center frequency of 430 MHz, then, the conversion between rotation
rate  $d \psi / d \nu$ and $RM$  is simply
\begin{equation}
{d\psi \over d\nu} 
 = -0.12953 \; {RM \over {\rm rad \; m}^{-2}} \; {\rm deg \over MHz};
\end{equation}
so we will use the two quantities interchangeably in what follows.

The linearly polarized position angle $\psi$ suffers a significant net
rotation across
the 32 frequency channels owing to interstellar and ionospheric Faraday 
rotation plus an instrumental position angle rotation term, respectively: 
\begin{equation}
{d\psi \over d\nu}|_{net}  =
{d\psi \over d\nu}|_{ISM}  +{d\psi \over d\nu}|_{Iono} +
{d\psi \over d\nu}|_{Instr} .
\end{equation}
In order 
to determine the interstellar 
component of the rotation and hence the interstellar magnetic 
field, and also to produce polarized pulse profiles not suffering significant 
linear depolarization, we next determined the net position angle rotation 
rate across the bandpass $d\psi / d\nu|_{net}$
and {\em{derotated}}  the linear polarization at each frequency by this amount.  

\begin{figure*}
\plotone{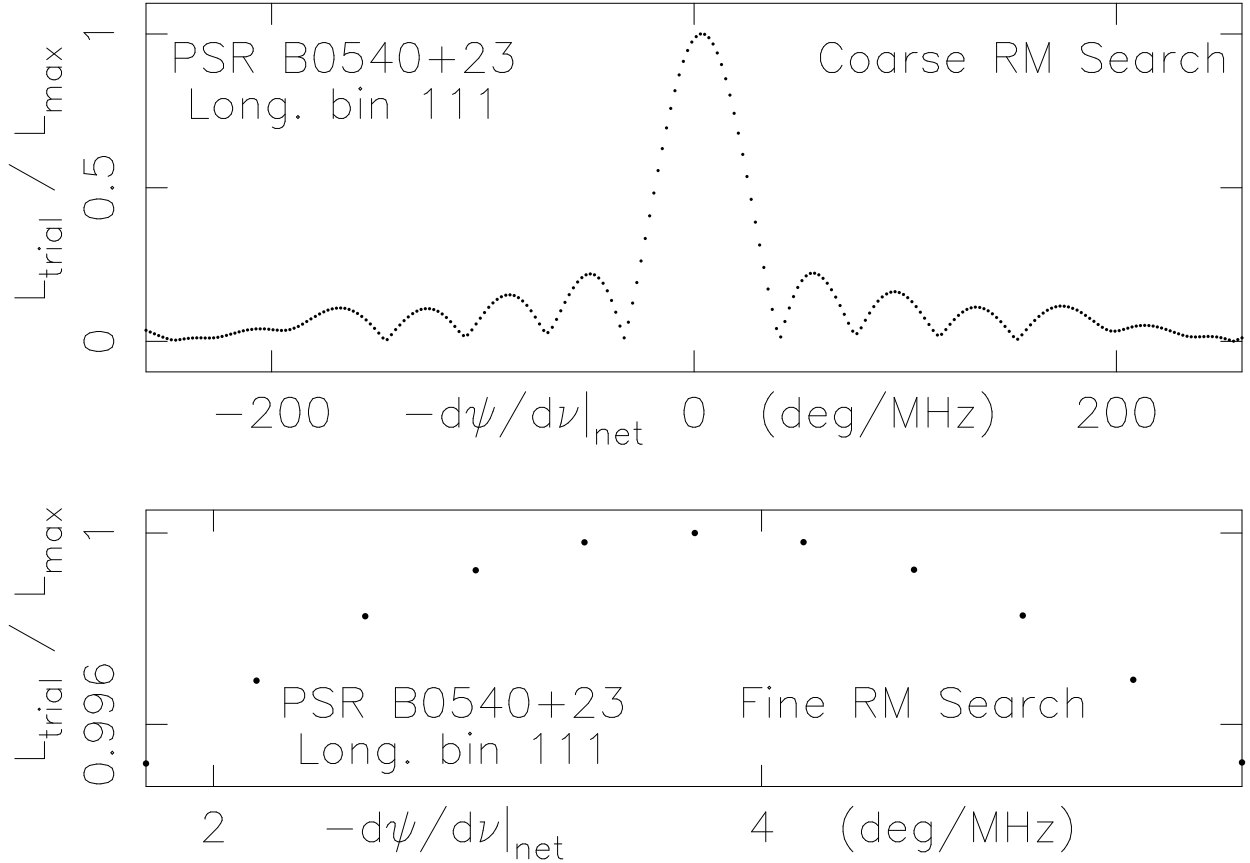}
\caption{$RM$ determination algorithm. For a given longitude bin 
of a two--minute 
observation, we calculate the linear polarization $L$ as a 
function of trial 
position angle derotation rate across our 32 frequency channels, 
$-d\psi/d\nu|_{net}$, 
in two passes.  First (top), $L$ is
calculated for 260 trial derotation rates corresponding to $RM$s 
between
$\pm 2000$ rad m$^{-2}$.  Next (bottom), we determine eleven 
trial $L$s at $RM$s
lying within $\pm 15$ rad m$^{-2}$ of the sweep rate that maximizes 
$L$ in the 
first step.  We then fit a parabola to the points. The
net sweep rate (and $RM$) adopted for this longitude bin correspond
to the sweep rate at the peak. In the displayed data, the adopted 
$-d\psi/d\nu|_{net} = 3.75$ deg / MHz.
\label{fig:Lvssweep}}
\end{figure*}

\begin{figure*}
\plotone{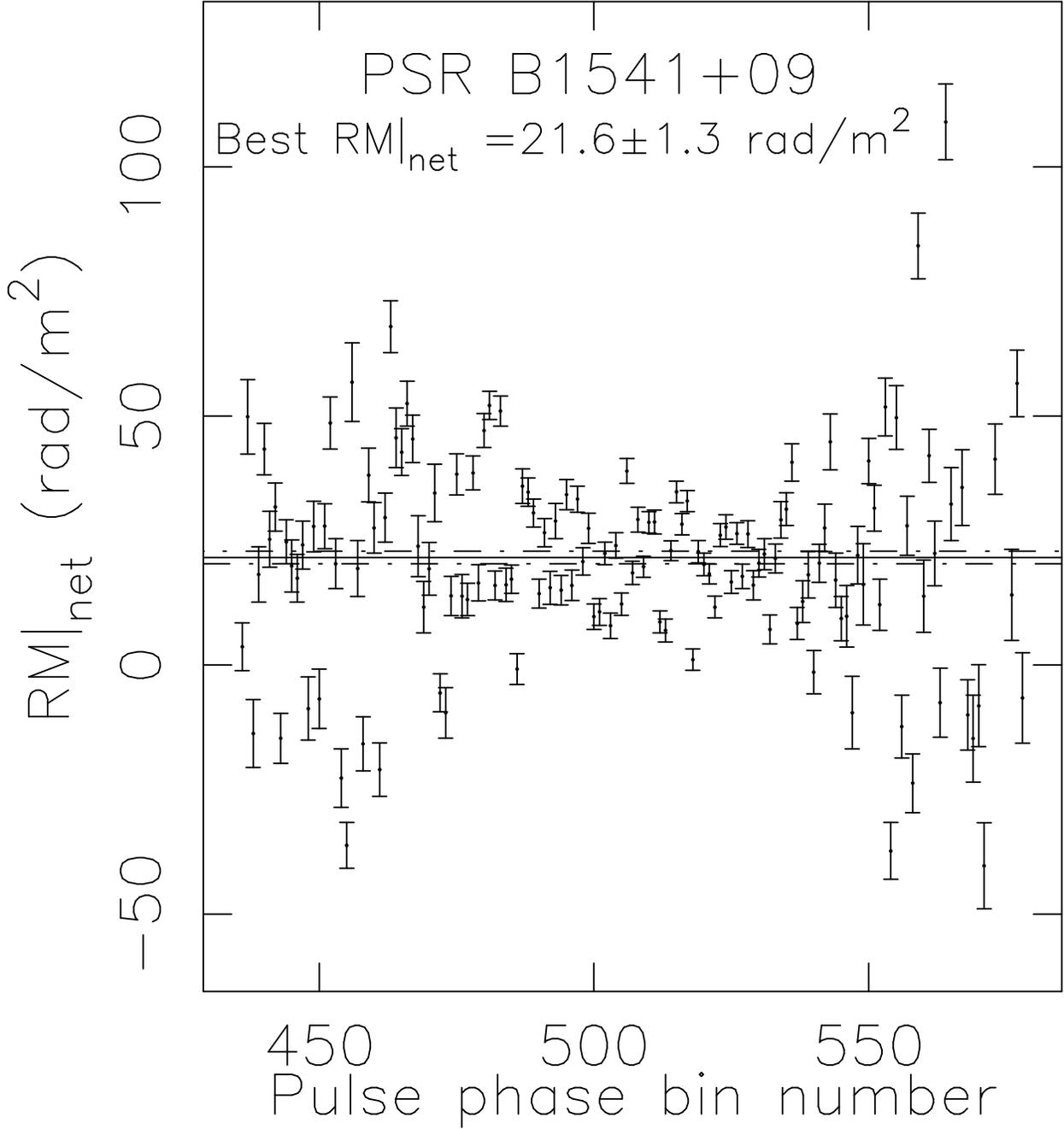}
\caption{Measured position angle sweep rate, $d\psi/d\nu|_{net}$, 
in $RM$ units, as a function of pulsar longitude for a two minute 
observation of PSR B1541+09.  Each data point represents the adopted 
$d\psi/d\nu|_{net}$ of a single pulse longitude bin (1/1024$^{th}$ of 
a full period) for the two minutes of observing, derived
as shown in Fig.~\ref{fig:Lvssweep}. The  error bars, used for weighting
individual points, are proportional to $\sigma_{\psi}$, and hence 
are large at longitudes  where there is little
linearly polarized power.  The solid horizontal line is 
the weighted average of the individual
points and the two dashed lines indicate the standard deviation of 
the mean (also reported at the top of the figure).
\label{fig:pavslong}}
\end{figure*}

To measure this position angle rotation rate with frequency, we performed two 
successively finer grid searches  at a given
longitude bin, as follows (see  Fig.~\ref{fig:Lvssweep}).
First  we chose each of 260 trial $\psi$ derotation rates corresponding to
$RM$s between $\pm2000$ rad m$^{-2}$, and then summed 
the derotated  linear Stokes parameters $Q$ and $U$ across the 32 frequency 
channels. Our
first estimate of net derotation rate of position angle with frequency  was 
the one that maximized $L$ in this first trial process.  Then we recentered 
the search on
this first estimate; and did 11 trial $\psi$ derotation
rates corresponding to $RM$s within $\pm15$ rad m$^{-2}$
of this value.  We then fitted a parabola to the peak of the latter
curve.
The adopted derotation rate of position angle with frequency, 
$-d\psi / d\nu|_{net}$,   was 
the one that maximized $L$ in this process.  The procedure was repeated at 
every longitude bin for which there was significant linearly polarized power 
(typically $L\ge 3\sigma_L$),
and a weighted average derotation for the two--minute scan, 
$<-d\psi / d\nu>|_{net}$, 
was then computed from the results at each longitude 
(see  Fig.~\ref{fig:pavslong}).  The uncertainty was 
taken to be the standard deviation of the mean.

To extract the interstellar Faraday rotation from the measured
 $<d\psi / d\nu>|_{net}$,
it is necessary to subtract the ionospheric and instrumental contributions
({\em{cf.}} Eqn. 8).  To
estimate the ionospheric Rotation Measure, we integrated a time--dependent 
model of the ionospheric electron density [International Reference Ionosphere
 1995; \citep{bili97}]
and geomagnetic field [International Geomagnetic Reference Field 1995;
\citep{bart97}]  through the ionosphere along the Arecibo--pulsar line of 
sight, yielding typical $RM|_{Iono}\sim (0.5 - 4)$ rad m$^{-2}$.  

The instrumental rotation rate across the bandpass was assumed to be constant 
during each given daily session, and was taken as the 
weighted difference between our measured Rotation Measures and the published 
values for six ``$RM$ calibrator pulsars.''  (The uncertainty in this quantity 
is the standard deviation of the mean difference.) Consequently our 
values depend on the correctness of previously published results and their 
constancy over time. Our calibrator list initially included every one of our
pulsars that additionally had a published $RM$ value.  It quickly became
clear that most of the published values were not correct at our observing
epoch, as they did not yield an instumental rotation consistent with other
pulsars.  We finally arrived at our final list of only six
``$RM$ calibrator pulsars'' (see Table 1) whose published $RM$'s could be
reconciled with our observations. 

After determining the ionospheric and instrumental rotation rates
as described above, we derived interstellar Rotation Measures  for each 
2--minute scan  for all of our pulsars via 
Eqn. 8.  Our final measured interstellar $RM$ for each pulsar was the 
weighted average of these results.

Comparison of our resulting ``measured'' interstellar
$RM$'s with the published values for the Calibrator pulsars constitutes a 
closed-loop  test of our calibration system, as we processed these  
pulsars  identically as the others we observed.  Note that  our ``measured'' 
values indeed agree with the published values for these pulsars to within the 
errors.  (See Table 1.) The many other pulsars'
published $RM$'s that disagree with our measurements will be discussed below.

\subsection{Final Polarized Pulse Profiles}

The weighted average position angle derotation $<-d\psi / d\nu>|_{net}$ 
was applied to the  Stokes parameter profiles 
at each of the 32 contiguous frequencies, and they were then summed to form a
polarized profile for the scan.  Each scan
on a given pulsar was crosscorrelated in two dimensions with a high $S/N$ 
full--Stokes profile template, in 
order to  shift it in time and position angle before accumulating it into a 
``grand average.''  Specifically, we first cross--correlated the total power 
$(I)$ scan and template as a function of longitude to obtain the time offset 
of each scan with respect to the template, and we rotated the scan's polarized 
profile in time by this amount.   Next, we crosscorrelated the linearly polarized 
$Q+iU$ scan and template profiles as a function of position angle offset to 
determine the best offset in $\psi$ between them, and we then rotated the 
scan's linearly polarized position angle by this amount.  Finally, the time--
and position angle--registered profile was accumulated into a grand average
waveform, as shown in Figs. 10--22.  We use the  IEEE convention for defining 
circular polarization \citep{kraus66}, in which left circularly polarized  
radiation is defined as that which is 
transmitted and received by a left-handed helix. The Stokes parameter 
$V = S_{\rm left} - S_{\rm right}$.

We estimate that in the worst cases, $V$ could be systematically in error
by $\pm5\%$ of $I$ due to the difficulty in calibrating left-- versus
right-- circularly polarized  radiometer channels via standard source
observations in crowded fields.  Additional possible errors in $V$ of up
to  $\pm0.15 L$ due to instrumental polarization were averaged down in 
most cases by multiple observations. 

These data were gathered in seven daily observing sessions from 1991 December 
through 1992 December.  The profile resolution is $1/1024$ of the pulse
period, or $0\fdg3516$.  For the weaker pulsars, the profile is 
boxcar--smoothed.  If such smoothing was done, the number of channels 
in the boxcar is  listed in the figure caption.

For those pulsars exhibiting little or no linear polarization, it is possible 
that an incorrectly  determined rotation measure has depolarized any   
intrinsic linear polarization. However, our rotation measure finding algorithm 
should have found the correct $RM$ if indeed there was sufficient $S/N$ in $L$.

\section{ROTATION MEASURES AND INTERSTELLAR MAGNETIC FIELDS}

In the following sections, we  present  our new $RM$ measurements.  We
then study both time variations in $RM$, and  the large--scale structure of 
galactic magnetic fields.

\subsection{Measured RMs and Time Variations}

Table 2 lists our rotation measures, previously published values (where
available), and the difference between them.  Note that many of our 
determinations differ from earlier measurements by more than the joint
uncertainties, suggesting that time variations are common.  
\citet{vO97} and \citet{H99} also list several pulsars having  $RM$s 
varying by several rad m$^{-2}$ on timescales of years.   

Eq. 3 shows that $RM$ variations might result either from  magnetic field
or electron density variations (or both).  It is possible to test for
the level of electron density variations if multiepoch dispersion measure 
($DM$) values are available.  While some 
pulsar  $DM$s  show very small time variations
(Phillips \& Wolszczan 1991; Backer et al 1993), others vary by several 
percent on these timescales 
\citep{DA93}.  \citet{vO97} 
found that $RM$s and $DM$s tended to vary in a correlated fashion,
while \citet{m74} detected a change in the $RM$ of PSR B0329+54 without
accompanying $DM$ variations. 

Fortunately, we are able to investigate possible sources for the $RM$
changes in our pulsars because \citet{HR03} measured $DM$s of many of 
them at 
approximately the same epoch as we gathered our $RM$ data.  Table 3
displays their and earlier $DM$ measurements on many of our $RM$ pulsars.  
It is 
clear from comparison  of the percentage changes in Tables 2 and 3 that 
$RM$s tend to vary far more on a several--year timescale than do $DM$s. 
Consequently it is evident that electron density variations are not
responsible for the bulk of the $RM$ changes.

Of course pulsars embedded in supernova remnants
such as Vela and the Crab Nebula, show large variations of both $RM$ and $DM$ 
(Hamilton et al 1985; Rankin et al 1988). The largest difference in Table 2 
between previously 
published and our measured $RM$s is for PSR J1532+2745=B1530+27, which 
does not lie in an SNR.  We have no other explanation for the discrepancy.
Multiple observations at several of our observing epochs confirm our $RM$ 
result on this pulsar. 

While it  appears from the above arguments that the temporal changes 
in $RM$ must  result from interstellar magnetic field variations,
caution is required.  Recent work by \citet{Ramach03} shows that large  
variations in $RM$ as a function of pulse phase are seen in some pulsars.  
(Our error analysis procedure accounts for these variations, so that our
tabulated values and uncertainties in $RM$s reflect this phenomenon.)
\citet{Ramach03} also  demonstrate that  this (spurious) 
apparent $RM$ variation with pulse phase can result from average--pulse
measurements of a pulsar having a quasi--orthogonal polarization mode 
emission process. Time variations in apparent $RM$ 
could then result from the balance of the two modes shifting slightly over 
time, rather than from interstellar magnetic field changes.

\subsection{The Galactic Magnetic Field in Selected Directions}

In the following sections, we use our new pulsar $RM$ data in conjunction
with previously measured values [summarized in \citet{m02} and supplemented by
recent measurements of \citet{MWKJ02}] 
to probe the galactic magnetic field structure
in selected directions.  Global analyses can be found in \citet{H99} and 
\citet{ID98}.  While  the random component of the field is comparable to
or even larger than the systematic structure \citep{OS93}; we focus here on the 
systematic part.  In order to investigate the
field as a function of location in the Galaxy, we use the NE2001 electron
density model \citep{CL02} and the pulsar dispersion measure $DM$ 
to determine the distance to each pulsar studied. In most cases, we select  
wedges of longitude along the galactic plane
for our studies, setting the longitude boundaries to isolate  global
magnetic field structures. (In all cases we have limited our studies to
 those pulsars with galactic latitudes $|b|<9\degr$).
In one case where a wedge cannot segregate the desired magnetic structure,
we instead select a region lying within a given distance range of a particular
spiral arm.   While the NE2001 model yields distances with typical 
uncertainties of 
some  tens of percent, the {\em{relative}} distances of various pulsars 
should be 
roughly correct within sufficiently small zones.  Since the rotation measure 
$RM$
is a path integral of the magnetic field,  a steady large--scale field will 
tend to appear as a constant {\em{slope}} (modulo changes in $n_e$) in 
plots of $RM$ versus distance, and a magnetic field reversal will lead to
a change in the  {\em{sign}} of the slope.  Note also from Eq. (1) that a 
positive (negative) $RM$ 
indicates that the radio signal is propagating parallel (antiparallel) to the 
field.

To set the overall context while invesigating particular zones, the reader
may examine a map of the galactic plane plotting all known low--latitude pulsar 
$RM$s  (see  Fig.~\ref{fig:globalfield}), with  superposed global field
directions deduced from this work.  This figure also displays
the spiral arms as presented by \citet{CL02}.

\subsubsection{Field Reversal Just Inside the Solar Circle}

It is difficult to analyze a magnetic field reversal when the reversal 
lies along a line oriented close to, but
not  parallel to, the line of sight.  Looking at pulsar $RM$s within 
a longitude wedge in this situation can be misleading because the 
reversal does 
not lie at a unique longitude.  Pulsars at similar longitudes and 
distances from the Sun, but on opposite sides of a reversal, may 
have drastically different $RM$s.  Consequently in this situation  we use 
different selection criteria than longitude.  It is useful 
to select pulsars within some given distance range from  the nearest
spiral arm in order to examine whether the magnetic field reversal follows 
the galactic spiral pattern.  With suitable adjustment of this distance 
window, we can 
see magnetic field effects for a very narrow and precise region of the 
interstellar medium near the spiral arms. For example we can investigate 
whether a reversal does indeed lie at a constant distance from a spiral arm.

The rotation measure depends upon four quantities, as seen from Eq. (1) --
distance, $n_e, B$, and the angle between the propagation vector and the 
magnetic field. Hence it is important 
to note that plots of $RM$ against distance cannot by themselves discriminate 
among the several factors that may affect $RM$.  This conclusion holds whether
one selects pulsars within either a wedge or a strip of the Galaxy. 

\begin{figure*}

\caption{\footnotesize $RM$s of low--latitude pulsars in the Galaxy, from \citet{m02},
\citet{MWKJ02}, and this work.  The horizontal dashed line
intersects the y--axis at the Sun, at coordinates $(0.0,8.5)$ kpc. The arrows
 indicate
schematically the direction of the field as ascertained from this  work (see
Figs. 4 -- 9 and text for further discussion). The  spiral arms are generated 
from routines provided by \citet{CL02}.}
\rotatebox{-90}{
\epsscale{0.6}
\plotone{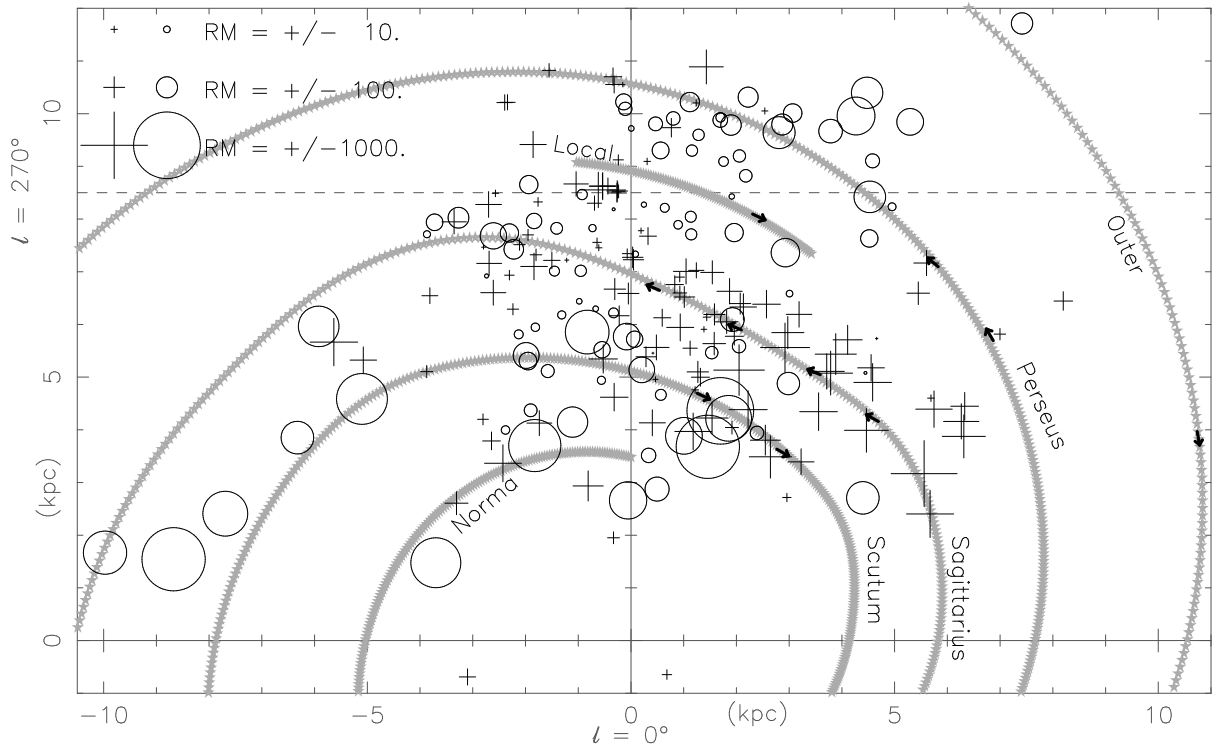}}
\label{fig:globalfield}
\end{figure*}

\begin{figure*}
\rotatebox{-90}{
\epsscale{0.7}
\plotone{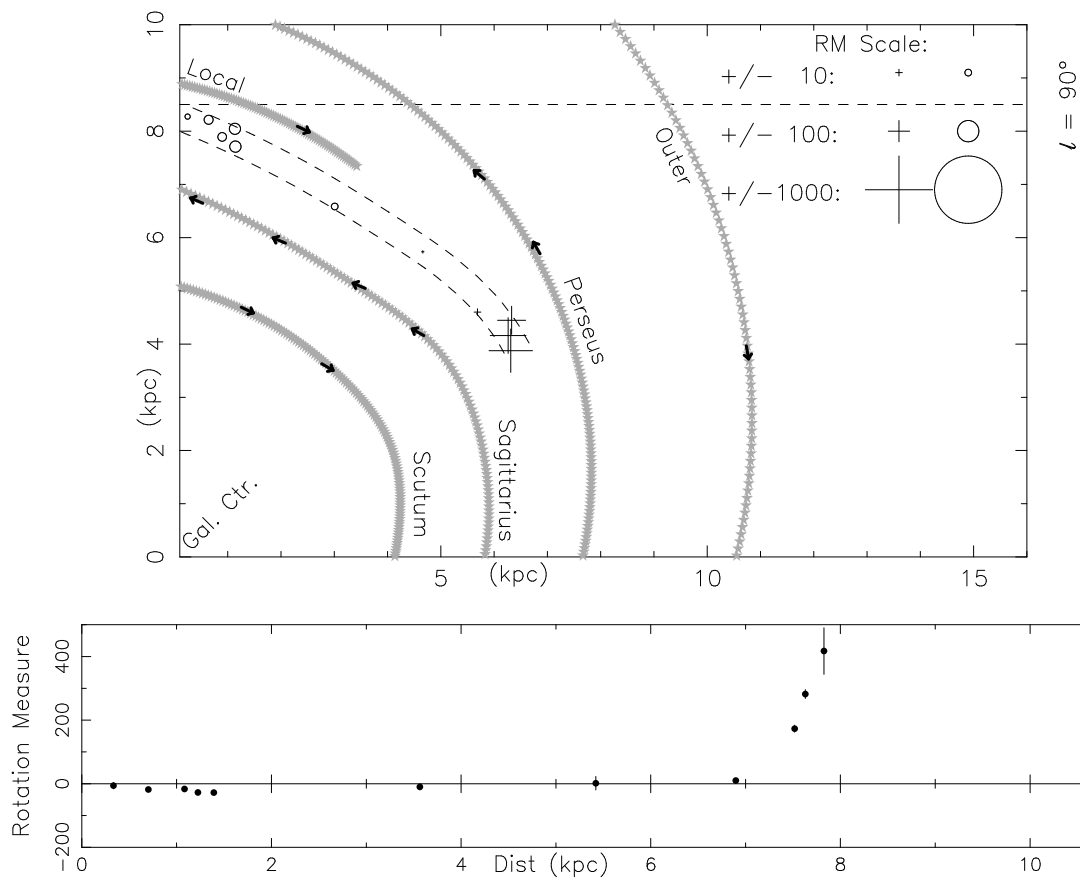}}
\caption{$RM$s of pulsars lying from 1.0 to 1.5 kpc outside the
Sagittarius arm (the first arm inside the Solar Circle), roughly in the 
$l\sim60\degr$ direction and near the galactic plane $(|b|<9\degr)$.  
Top:  The selected region is delineated
by curving dashed lines.  See Fig. 3 caption for additional details.  Bottom: 
$RM$ versus distance in the selected
region.  Note that there is a clear null in the magnetic field all 
the way from the solar vicinity out to $d\sim 7$ kpc.
\label{fig:atopreversal}}
\end{figure*}

The field reversal thought to exist several hundred pc inside the Solar
 Circle (Thomson \& Nelson 1980;  Han \& Qiao 1994; Rand \& Lyne 1994) 
is such a case. The field is clockwise near the Sun when viewed from
above;  inside the 
reversal it changes to counterclockwise from above (see 
Fig.~\ref{fig:globalfield}).  \citet{RL94}  found negligible $RM$s
throughout the wedge $60 < l < 80\degr$ for all pulsars within about 5 kpc
of the Sun, and attributed the null to a large--scale field reversal.  Their
 result is no longer supportable with the larger number of
measurements accumulated since then. In fact we find that there is {\em{no}}
unique {\em{longitude}} range in this vicinity that traces the null for great 
distances.  However,   for the {\em{strip}} lying approximately midway 
between the Local arm and the Sagittarius (next inner) arm toward $l\sim60\degr$-- 
specifically from (1.0--1.5) kpc outside the Sagittarius arm  
(see  Fig.~\ref{fig:atopreversal}), we find that $RM$s cluster near zero for 
the first 7 kpc from the Sun.    Because these pulsars are  at varying 
distances from the Sun (from 0.3 to nearly 7 kpc), we must be looking 
along an actual null in the magnetic field extending for $\sim7$ kpc.  In 
this case, instead of surmising the existence 
of a reversal on the basis of  {\em{trends}} in the sign of $RM$, we clearly 
see the reversal itself.  We also find that if we widen the zone much beyond
our current limits, the $RM$s tend systematically away from zero as expected
on opposite sides of a reversal. (See below for details.) Therefore it appears
that the null is confined to a width of less than 0.5 kpc  over
the 7 kpc range.  It is important to emphasize that this null, whose location
is now by far the best--known for any large--scale field reversal in the 
Galaxy, clearly lies midway {\em{between}} arms rather than along one.

The rapid, systematic rise in $RM$ for $d > 7$ kpc signals that
the null no longer lies $\sim1$ kpc outside the Sagittarius arm here.  
Unfortunately, there are  not enough pulsars with measured $RM$s near and
beyond here to map out exactly what is happening in this vicinity, but we note
that the Local arm is presumed to end  and the Perseus (next outer)
arm also approaches closer to the Sagittarius arm near here. Presumably these 
global galactic structures are perturbing the well--behaved pattern that we
were able to trace out for a considerable distance up to this location. $RM$
measurements of additional, more distant pulsars in this direction will help
to clarify the situation.

We have also investigated the local reversal in the opposite direction, i.e.,
toward $l\sim270\degr$.  The reversal is not nearly as well--defined, but
in general the trend is consistent with what is found toward $l\sim60\degr$
(see  Fig.~\ref{fig:globalfield}): The local field reversal midway between
the  Local arm and the next arm interior to it (Sagittarius) separates clockwise
fields near the Local arm from counterclockwise fields near the next interior 
one; causing predominantly positive $RM$s in the former case and negative
ones in the latter.  The few counterexamples probably result from
Faraday screens along particular lines of sight, a point emphasized by
\citet{MWKJ02}.  The overall field
structure at large distances toward $l\sim280\degr$ is unknown because
of perturbations caused by the Gum Nebula.

\subsubsection{Field Structure Interior to the Nearby Field Reversal}

\begin{figure*}
\rotatebox{-90}{
\epsscale{0.7}
\plotone{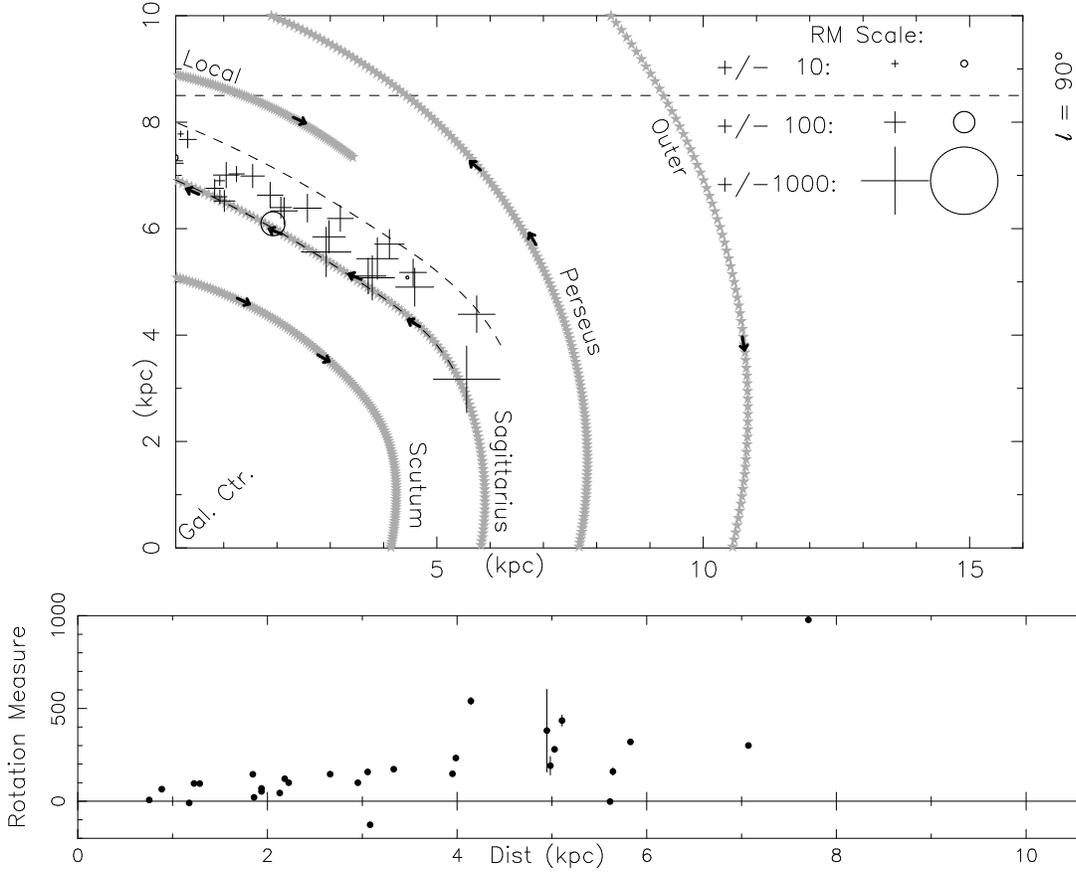}}
\caption{$RM$s of pulsars lying in the region between the Sagittarious 
arm and 1.0 kpc exterior to it near the galactic 
plane $(|b|<9\degr)$.  Top:  The selected region is delineated
by  curving dashed lines. See Fig. 3 caption for additional details.
Bottom:  $RM$ versus distance in the selected region.   Note that a
constant magnetic field would lead to a constant {\em{slope}} (modulo
variations in $n_e$) in this plot, since $RM$ is a path integral quantity.
\label{fig:stripeminus1}}
\end{figure*}

\begin{figure*}
\rotatebox{-90}{
\epsscale{0.7}
\plotone{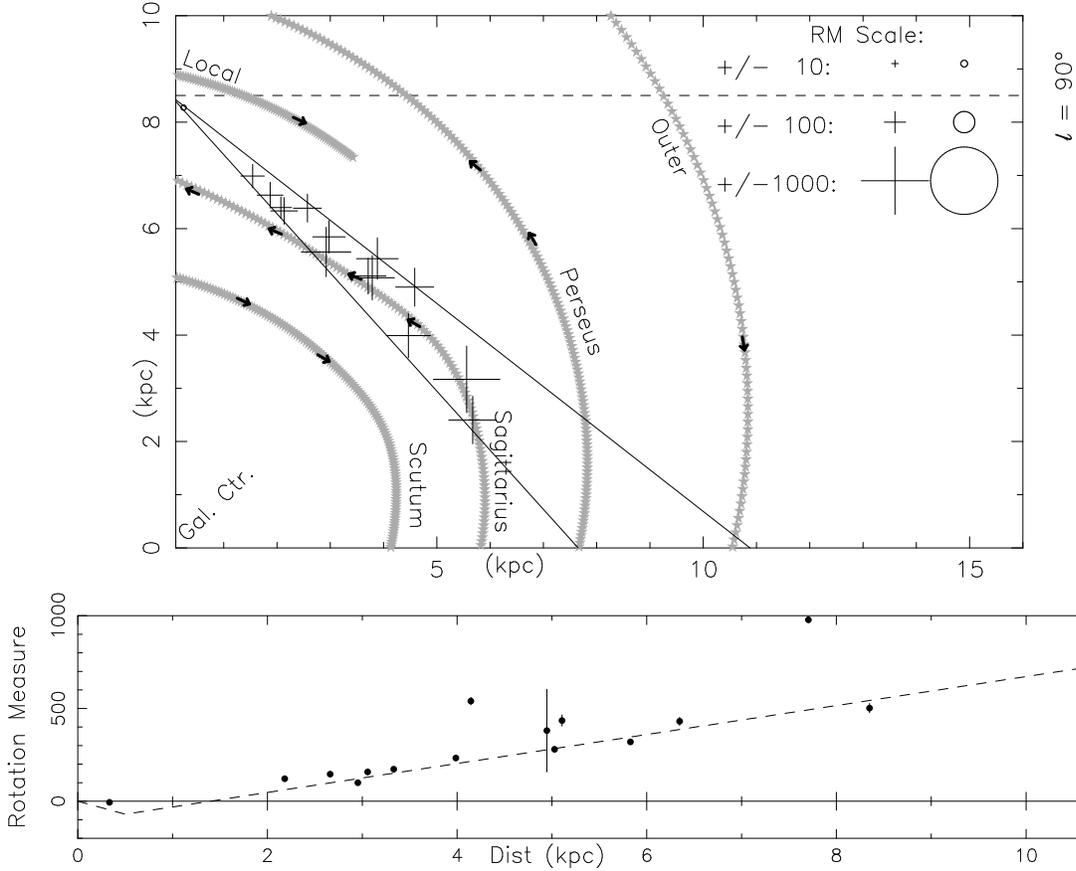}}
\caption{$RM$s of pulsars lying in the range $42<l<52\degr$ near the galactic 
plane $(|b|<9\degr)$.  Top:  The selected region is delineated
by a wedge. See Fig. 3 caption for additional details. Bottom:  $RM$ versus 
distance in the selected
region.  Most extragalactic sources in this direction have $RM\sim500$ rad
m$^{-2}$ \citep{C92}.  The dashed line illustrates the trend of increasing
$RM$ with distance due to a roughly uniform counterclockwise field along 
the outer edge of the Sagittarius arm.
\label{fig:42to52}}
\end{figure*}

We next investigate the field lying in three 1--kpc wide strips interior to the 
nearby field reversal. The first segment selects pulsars from the reversal 
inward to
the Sagittarius arm. (See  Fig.~\ref{fig:stripeminus1}.) Note the clear trend
of increasing $RM$ with distance, which indicates that the field is relatively 
steady in this region.  The narrow wedge from $42\degr < l < 52\degr$, whose 
inner edge cuts just inside the Sagittarius arm (see  Fig.~\ref{fig:42to52}),
also illustrates that the field from the local reversal in to Sagittarius
arm is approximately steady, as there is again a regular trend of increasing 
$RM$ with distance. Most extragalactic sources in this direction
exhibit $RM\sim500$ rad m$^{-2}$ \citep{C92}, so there is no evidence
for additional major reversals beyond the pulsars.

\begin{figure*}
\rotatebox{-90}{
\epsscale{0.7}
\plotone{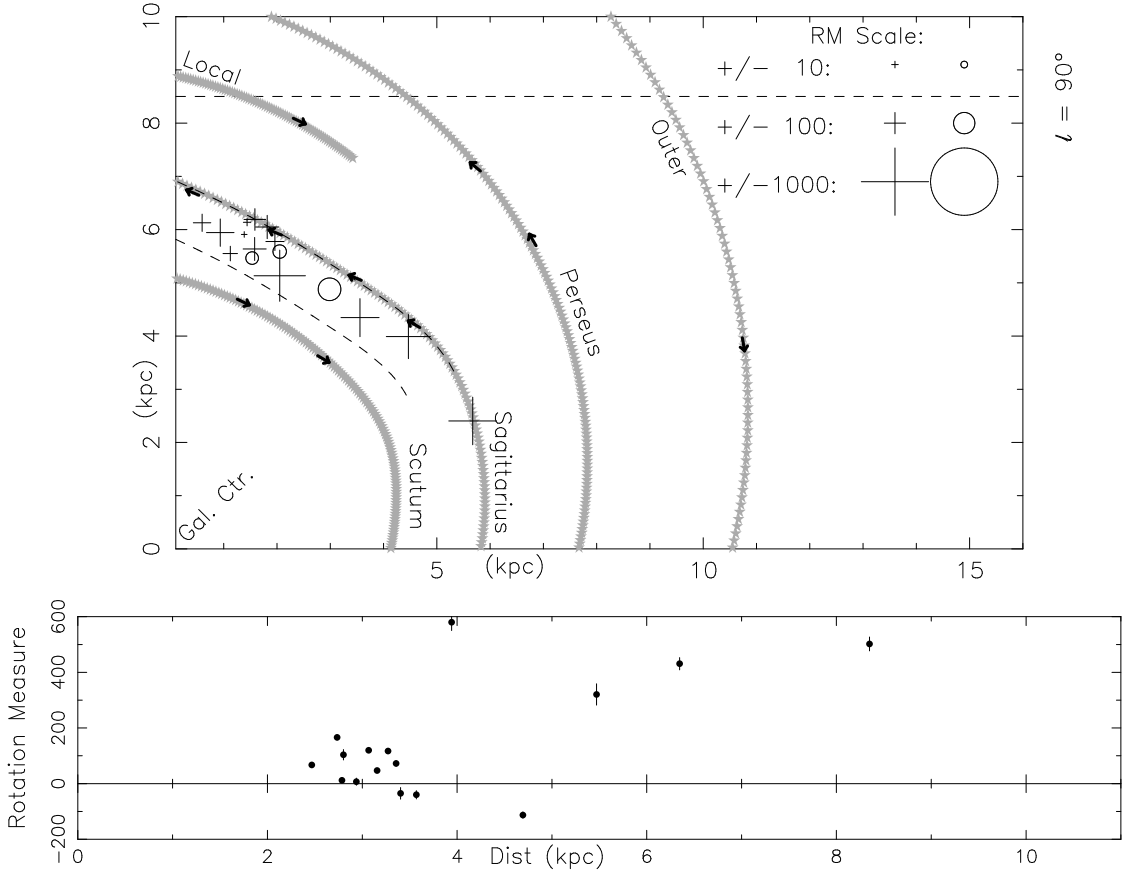}}
\caption{$RM$s of pulsars lying in the region between the Sagittarious 
arm and 1.0 kpc interior to it near the galactic 
plane $(|b|<9\degr)$.  Top:  The selected region is delineated
by curving dashed lines. See Fig. 3 caption for additional details. Bottom:  
$RM$ versus distance in the selected region.
\label{fig:stripeminus2}}
\end{figure*}

The second strip interior to the local field reversal selects pulsars from 
the Sagittarius arm to 1 kpc on its 
interior side. (See  Fig.~\ref{fig:stripeminus2}.) The general trend of
 $RM$ with distance in this strip is similar to that in the first region. 
As $RM$ is a path integral quantity, this similarity suggests that 
a second low--field region (i.e., a second reversal) is being traversed  somewhat 
interior to  the Sagittarius arm.  The farther pulsars in these two strips
tend to  show {\em{higher}} $RM$s because  their signals propagate more nearly
parallel to the field.

\begin{figure*}
\rotatebox{-90}{
\epsscale{0.7}
\plotone{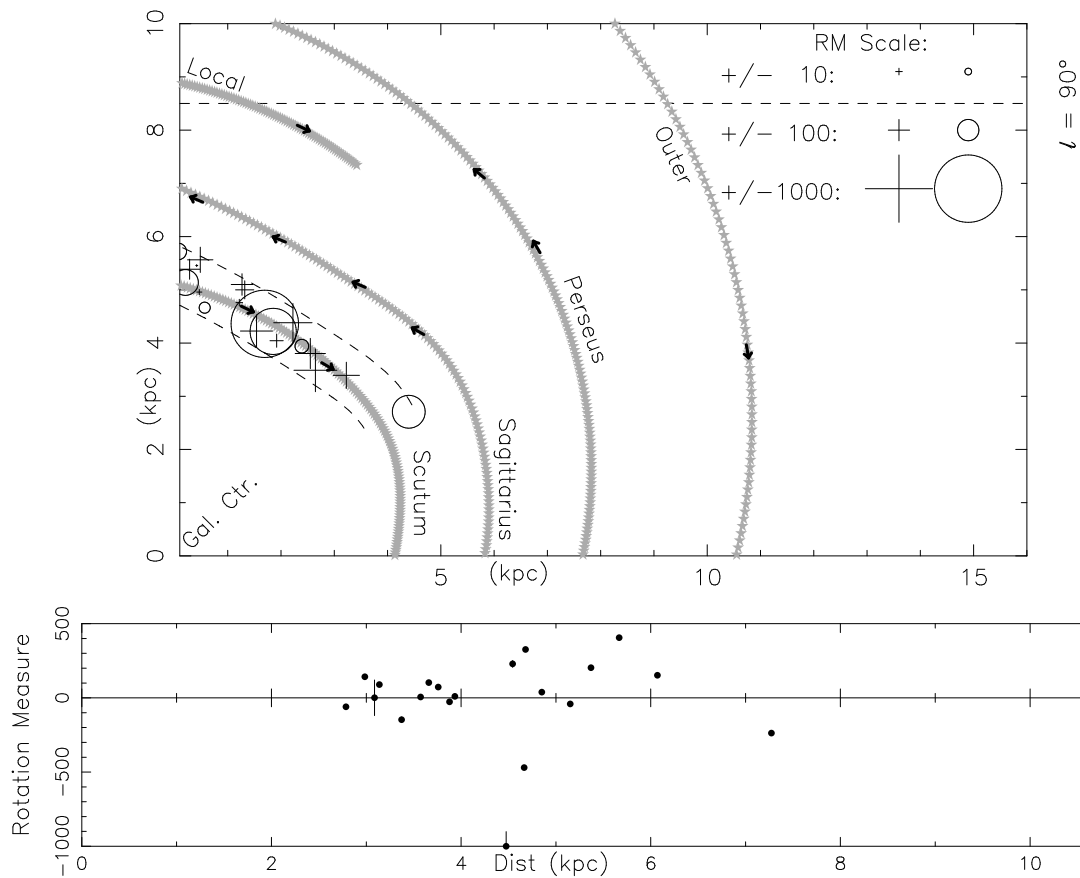}}
\caption{$RM$s of pulsars lying in the region between 1 and 2 kpc interior to
 the Sagittarious arm, roughly atop the Scutum arm, near the galactic 
plane $(|b|<9\degr)$.  Top:  The selected region is delineated
by curving dashed lines. See Fig. 3 caption for additional details.  Bottom:  
$RM$ versus distance in the selected region.
\label{fig:stripeminus3}}
\end{figure*}

\begin{figure*}
\rotatebox{-90}{
\epsscale{0.7}
\plotone{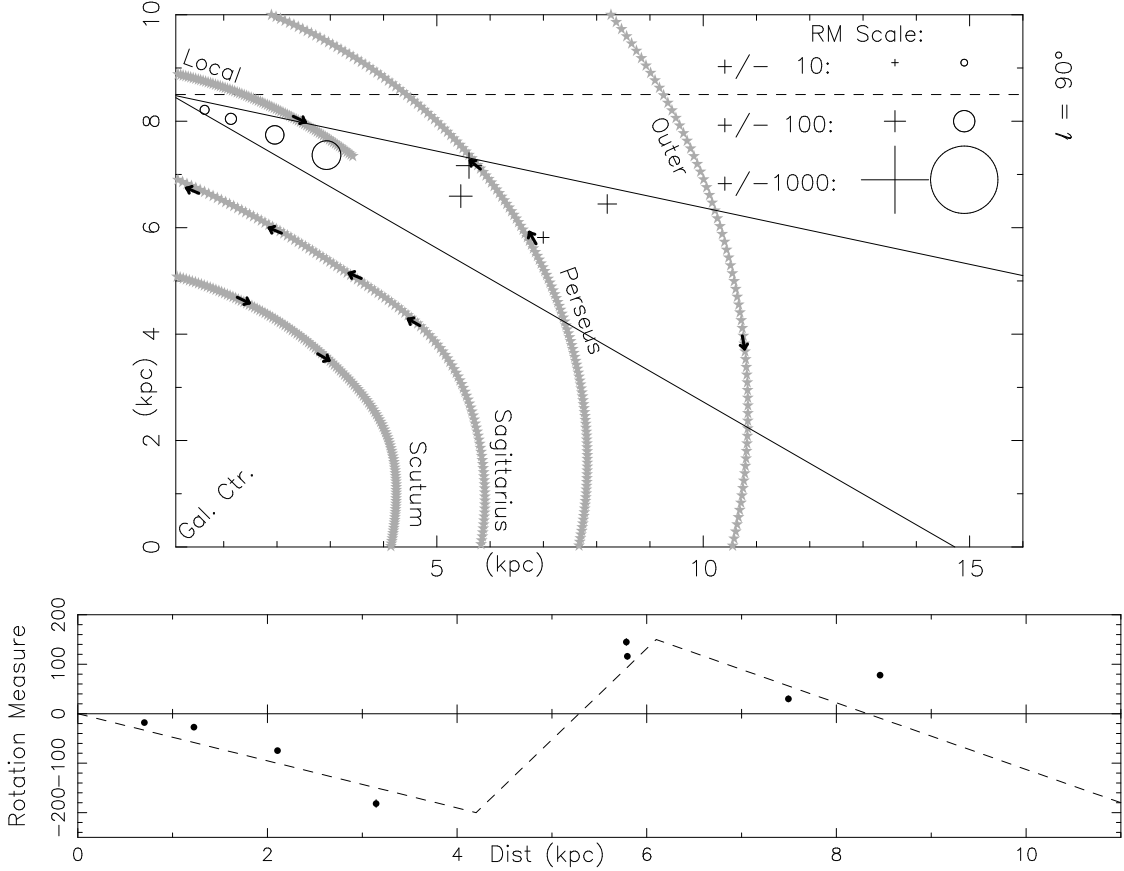}}
\caption{$RM$s of pulsars lying in the range $60<l<78\degr$ near the galactic 
plane $(|b|<9\degr)$.  Top:  The selected region is delineated
by a wedge. See Fig. 3 caption for additional details. Bottom:  
$RM$ versus distance in the selected
region.  Extragalactic Source $RM$s in this direction are negative 
\citep{C92}.   The dashed line illustrates schematically the trends in
$RM$ versus $d$, suggesting field reversals before and after the Perseus arm.
(Note that a field reversal leads to a change in the sign of the {\em{slope}} 
of $RM$ versus distance, since $RM$ is a path integral quantity.)
\label{fig:60to78}}
\end{figure*}

 The third strip interior to the local reversal, (see  
Fig.~\ref{fig:stripeminus3}), which encompasses the
Scutum arm, shows  more negative $RM$s than the first and second strips, especially
at larger distances. This indicates that
these pulsars' signals begin their journey  through a sufficiently long 
region of clockwise 
field near the pulsar, before reaching the second reversal described above,  
to approximately 
nullify the opposite field along much of the rest of the path.  (See the arrows
on Fig.~\ref{fig:60to78}, which sketch  the deduced field directions).

It is interesting to note that our work suggests that both of the nulls/reversals 
lie between arms, while \citet{ID98}, \citet{H99}, and \citet{H02} indicate 
{\em{maximal}} field magnitude in these regions.  Much of the discrepancy
can be attributed to the new NE2001 electron density model bringing most pulsars 
closer to Earth, while some of the change is due to adjustments in the location of 
the spiral arms themselves.    [\citet{H02} find yet another (third) 
reversal even farther in, also near an arm -- the Norma arm. 
The latter $RM$s have not yet been published so we are not able to study them 
with the aid of the new NE2001 model -- determined distances.]

\subsubsection{Field Reversals External to the Solar Circle}

We next investigate the longitude range $l\sim(60-78)\degr$
[see  Fig.~\ref{fig:60to78}].
The lower longitude limit of this wedge ensures that lines of sight remain
exterior to the local reversal which lies just inside to the Solar Circle.  
The
nearest several kpc of the wedge skim near or along the Local arm.
  The $RM$s near the Sun grow more  
negative with distance out to $d\sim3$ kpc, as  expected from a clockwise 
local magnetic field.
 Yet in the range 6 $\lesssim$ $d$ $\lesssim$ 8 kpc, $RM$s are positive; while
{\em{extragalactic}} $RM$s are again negative  
\citep{C92}.  Consequently {\em{two}} field reversals, each 
corresponding to a change of
slope in $RM$ versus distance, are required at $d\sim4.5\pm1$ kpc and 
$d\gtrsim6$ kpc, as indicated
schematically on the Figure.  The nearer reversal is located interior to the 
Perseus (next outer) arm; the farther reversal somewhere beyond it. As
in the interior Galaxy, we find that reversals/nulls tend to occur between arms.
(The field between these two exterior reversals, near the Perseus arm, is 
counterclockwise [see Fig.~\ref{fig:globalfield}].)  \citet{H99}also suggested 
that there may be two reversals exterior to the Sun; 
our contribution rests in providing several new measurements and 
in the application of the NE2001 model to estimate distances, both of which 
refine the arguments for two exterior reversals, at least in this direction.  
Note however that there is no clear evidence for reversals toward or beyond 
the Perseus arm at higher longitudes (Brown \& Taylor 2001; Mitra et al 2002).

\section {POLARIZED PROFILES AND MORPHOLOGICAL CLASSIFICATIONS}

Figures 10 -- 22 show the full polarization profile for each pulsar. (In a few 
instances, 
linear $(L)$ and/or circular $(V)$ polarization is not displayed because 
it is too weak.)   We discuss each profile and its classification below. The 
classifications are based on the Rankin (1983) morphological model, which has 
recently been further elucidated (Mitra \& Rankin 2002; Rankin \& Ramachandran 2003).

\subsection{PSR B0045+33 = J0048+3412; ~~~~P = 1\fs217;~~~~~~~~Fig.~10}

\citet{GL98} show this pulsar at 408, 606, and 1420 MHz.  Our 430 MHz profile is
similar to the first of these, although with higher $S/N$.  There is significant
 negative circular, and  little linear polarization. \citet{KL99}
show a similar profile in total power at 102.5 MHz.  This appears to be a core 
single ($S_t$) based on its frequency evolution and  circular polarization.

\subsection{PSR B0301+19 = J0304+1932; ~~~~P = 1\fs388;~~~~~~~~Fig.~10}

\citet{W99} cite extensive multifrequency observations on this pulsar and 
classify it as a classic conal double $(D)$. \citet{EW01} successfully fitted a 
rotating vector model (RVM) to the 21 cm position angle data, further 
supporting a conal classification since Rankin's model indicates that 
only conal emission 
is expected to exhibit the RVM. \citet{KL99} show that a central component 
appears at 102.5 MHz, as
expected of a core.  The circular polarization in our measurements and those of 
\citet{GL98} switches sign toward the latter of
the two principal components.  The \citet{R83} 430 MHz Arecibo observations 
show positive circular polarization throughout the profile, as do
measurements at higher frequencies starting at 610 MHz \citep{GL98}. It is
unclear whether the changes are due to instrumental
polarization or temporal variations.  \citet{Rama02} note that some of the 
early Arecibo polarimetry
was flawed by the use of Gaussian--shaped, rather than square, filters.

\subsection{PSR B0523+11 = J0525+1115; ~~~~P = 0\fs354;~~~~~~~~Fig.~10}

\citet{RSW89} suggested that  the sign--changing circular polarization under
 the leading components at 21 cm indicated a possibly merged leading cone 
and core. 
This  circular polarization signature is also seen at 1.6 GHz \citep{GL98}.  
However, our 
430 MHz profile shows  essentially {\em{no}} circular polarization under the 
leading  (and trailing) components, and only positive $V$ through the middle 
regions.  Our linear polarization is significantly higher under the bridge 
than is seen at 21 cm.  The position angle sweeps down under the bridge, and 
then jumps discontinuously at a null in $L$, finally settling at a constant 
value in a small trailing component that is also barely evident at 21 cm 
\citep{W99}.  According to the Rankin model, the central position angle 
traverse results from 
conal emission, which would indicate that our line of sight is tangent to 
the edge of a cone there.  Then the leading and (weak) trailing components 
could be emission
from a second cone, although their linear polarization properties are not
consonant with expected conal emission. \citet{KL99} show two principal 
components separated by $\sim8\degr$ at 102.5 MHz, {\em{less}} than at higher
frequencies. Consequently, the pulsar now fits less clearly into any Rankin  
class.

\subsection{PSR B0525+21 = J0528+2200; ~~~~P = 3\fs746;~~~~~~~~Fig.~10}

Our profile differs from the 430 MHz profile of \citet{R83} in that we see 
substantially more circular polarization under the second peak.  
\citet{W99} note that the central bridge has some core--like properties, but 
they still support the \citet{RSW89} double $(D)$ classification. The sucessful 
RVM fit of \citet{EW01} further supports this classification.

\subsection{PSR B0540+23 = J0543+2329; ~~~~P = 0\fs246;~~~~~~~~Fig.~11}

Most earlier work [summarized in \citet{W99}] suggests that this is a core 
single $(S_t)$. Our profile has the opposite sign of circular polarization as do
\citet{RB81} at the same frequency.  Recent 408 MHz polarized profiles by 
\citet{GL98} are similar to ours, so it appears that the sign of circular
polarization is the same up through at least 10 GHz.

\subsection{PSR B0609+37 = J0612+3721; ~~~~P = 0\fs298;~~~~~~~~Fig.~11}

Our linear position angle shows three well--defined values
under what appear to be three different pulse components.  There is no
evidence of an RVM--style position angle sweep.  The \citet{GL98} 1.4 GHz
linear profile shows a similar form in linear polarization and position angle 
sweep, though with larger amplitude in $L$.  [\citet{W99} were unable to
measure $L$ in this pulsar at 1.4 GHz.]

 The \citet{GL98} 410 MHz 
profile shows significantly more negative circular polarization
past the profile peak than does ours.  It is not clear if this discrepancy is 
due to calibration errors or to temporal changes in one  of the profiles. 
(Both profiles were measured roughly contemporaneously.)  However,
it is interesting to note that the multifrequency 
circular measurements of \citet{GL98} 
appear self--consistent up through 1.4 GHz, yet our  measurements indicating
negligible circular polarization
also appear roughly self--consistent with our contemporaneous 1.4 GHz 
observations \citep{W99}.   

The best profiles  at 0.4 and 1.4 GHz exhibit  three distinct linearly 
polarized components, edge depolarization,
and hints of sign--changing circular polarization under the (not very 
prominent) central component,
 arguing for a triple $(T)$ classification.  At 102 MHz, \citet{KL99} show 
two principal components separated by $\sim20\degr$, with a possible third
trailing one at a similar separation.  These data support the $T$ 
classification, as the outermost (conal) components have spread and the
central (core) component dominates, as expected at low frequency.

\subsection{PSR B0611+22 = J0614+2229; ~~~~P = 0\fs335;~~~~~~~~Fig.~11}

This pulsar is clearly a core single $(S_t)$.  See \citet{W99} for a summary 
of multifrequency observations. New 102.5 MHz observations by \citet{KL99}
also display a single form, although the resolution is rather coarse. The 
current observations show polarized profiles 
similar to the 0.4 GHz measurements of \citet {RB81} and \citet{GL98}.
It is interesting to note that the circular polarization feature migrates to
later longitudes at higher frequencies.  

\subsection{PSR B0626+24 = J0629+2415; ~~~~P = 0\fs476;~~~~~~~~Fig.~11}

\citet{W99} summarize measurements at other frequencies and conclude that the 
classification is a core single $(S_t)$, possibly with a merged core and 
leading conal component. Our observations and those of \citet{GL98} represent 
the first 
0.4 GHz polarimetry on this source. The current measurements display
polarization quite similar to higher frequencies.  Note particularly the
very high fractional $L$ and sign--changing $V$ in the leading part of the 
profile. We reclassify this pulsar as Triple $(T)$, while concurring with
the earlier suggestion of merged leading components. 

Surprisingly,
102 MHz measurements by \citet{KL99} do not verify the profile broadening
measured by \citet{IMS89}. This new observation does not rule out
 conal components, however.
While the original \citet{R83} classification
scheme indicated that low frequency profile broadening was a general 
property of conal emission, recent work by \citet{MR02} shows that 
{\em{inner}} conal pair spacing remains roughly constant with frequency.

The reclassification is not unexpected.  Earlier efforts had indeed
noted the three components, but the lack of 0.4 GHz polarimetry at that
time made it less apparent that all were sufficiently distinct at this
frequency, where the classification is supposed to originate.  

\subsection{PSR B0656+14 = J0659+1414; ~~~~P = 0\fs385;~~~~~~~~Fig.~12}

The profile shows almost $100\%$  fractional $L$ and moderate $V$ 
over  longitudes between $\sim\pm20\degr$, strongly indicating core emission. 
There is low--level, less polarized, emission before and after this component, 
which was first identified at 1.4 GHz (Rankin et al 1989; Weisberg et al. 1999)
and ascribed to possible conal emission.  However, as pointed out by
\citet{W99}, these components are not prominent at $\sim 5$ GHz as expected of 
conal emission,  
although the noise level is rather high  [\citet{I94}, 
\citet{seiradakis95}].The position angle in the skirts appears to smoothly follow 
the central component's, also suggesting that they are part of it. 
The  profile is also single at 102.5 MHz \citep{KL99}.
Consequently our classification remains core single $(S_t)$.

\subsection{PSR B0751+32 = J0754+3231; ~~~~P = 1\fs442;~~~~~~~~Fig.~12}

Our 430 MHz profile is similar to the 21 cm polarized profiles of \citet{RSW89}
and \citet{W99}.  There is a classic S--shaped position angle swing and a leading 
orthogonal mode jump in both.  The circular polarization $V$ is
somewhat lower at 430 MHz, and the leading notch of linear power at 1400 MHz
has declined significantly.  Our data support the earlier conclusions that there
is no evidence for core emission, and we support the conal double $(D)$
classification.

\subsection{PSR B1133+16 = J1136+1551;~~~~P = 1\fs188;~~~~~~~~Fig.~12}

Our 430 MHz linearly polarized profile is significantly different than 
the 408 MHz profile of \citet{GL98}, in that the relative intensity of the 
two primary 
peaks is reversed.  As our observations are roughly contemporaneous with theirs,
the variation probably represents a short--term change.  The $V$ profiles are
also somewhat different, but this may represent the effects of instrumental
polarization.

We see measurable linearly polarized emission out to $-10 \degr$ longitude,
well down on the leading skirt.  This faint emission was first discovered 
in the 21 cm observations of \citet{W99}. 

As summarized in \citet{W99}, most evidence points to a double ($D$) 
classification.  Our sign--changing circular and high linear polarization in
the saddle might argue for a central core.  However, the smooth, S--shaped
position angle swing seen here and in single pulse 21 cm observations [\citet{s84}; 
\citet{ganga99}] argues for conal emission alone.

\subsection{PSR B1237+25 = J1239+2453;~~~~P = 1\fs382;~~~~~~~~Fig.~12}

Multifrequency polarimetry and classification are summarized in \citet{W99}.   
Our 430 MHz profile agrees with the 408 MHz profile of \citet{GL98}, except 
that our finer resolution
enables us to distinguish the inner and outer conal emission much better.  
The
430 MHz polarimetry of \citet{R83} is similar except that her $V$ is 
significantly 
different.  This is a classic multiple $(M)$  profile.  Note that the 
very high 
$L$ in the second component is unusual for a cone and may suggest
blended core and conal emission.  The third component, however, has all the 
markings of a core, including  sign--changing circular polarization and
particular prominence at 200 MHz \citep{R83}.

\subsection{PSR B1530+27 = J1532+2745;~~~~P = 1\fs124;~~~~~~~~Fig.~13}

We have measured the Stokes parameters of the postcursor [discovered by 
\citet{W81}] for 
the first time.  It is virtually $100\%$ linearly polarized, with a rather 
constant 
position angle that contrasts with the sweep of the main components' $\psi$. 
 Our  profile has better $S/N$ than \citet{BCW91}, which 
also represents Arecibo 430 MHz polarimetry.  The 408 MHz profile of \citet{GL98} 
is similar to our main pulse profile, although its $V$ is closer to zero.  
Careful analysis of our individual scans suggests that our $V$ may be 
influenced by instrumental 
polarization, so that the \citet{GL98} $V$ may be superior. We agree with 
earlier conal single $(S_d)$ classifications, especially since the profile
exhibits edge depolarization and becomes clearly double at lower frequencies.  
The nature of the postcursor emission remains unclear.

\subsection{PSR B1541+09 = J1543+0929;~~~~P = 0\fs749;~~~~~~~~Fig.~13}

\citet{W99} observed this pulsar at 21 cm and summarized extensive multifrequency
 observations.
\citet{EW01} successfully fitted an $RVM$ curve to 21 cm position angle
data after unweighting an anomalous central region.
Our 430 MHz polarimetry is similar to that of \citet{R83}, except that  she 
found higher fractional $L$ in the two outer, conal components.  We also note 
that our total power component at $-25 \degr$ longitude appears more distinct 
than in 430 MHz Arecibo observations of \citet{R83} and \citet{HR86}.  We 
confirm the earlier $T$ classifications.

\subsection{PSR B1612+07 = J1614+0737;~~~~P = 1\fs206;~~~~~~~~Fig.~13}

\citet{W99} presented 21 cm polarimetry and summarized observations from 100 MHz
to 5 GHz.  They found that it was not possible to unambiguously classify this 
pulsar, and called for polarimetry at more frequencies.  Our 430 MHz profile
shows a systematically sweeping position angle, unlike the rather flat curve
at 21 cm.  According to classification criteria of the \citet{R83} model, the
position angle sweep suggests conal emission, adding support to a tentative
classification as conal single $(S_d)$.  We emphasize that the pulsar still
does not fit neatly into this classification, however, because of its behavior
at very low and very high frequencies.

\subsection{PSR B1737+13 = J1740+1311;~~~~P = 0\fs803;~~~~~~~~Fig.~14}

The multifrequency properties of this multiple $(M)$ pulsar between 100 MHz 
and 5 GHz are summarized in \citet{W99}. With few exceptions, this pulsar
matches the canonical behavior predicted for core and conal emission by 
\citet{R83}, including  conal spreading and edge depolarization, 
and a prominent core at low frequencies.  Based on its spectral behavior, 
our component at $0 \degr$ longitude is clearly a core: It is the most 
prominent component at low frequencies but comparable to other components 
at  higher frequencies up to 2.3 GHz \citep{HR86}. There are two conal components 
on each side of it. In our profile, the first trailing conal component is 
barely distinguishable from  the core in total power, but is clearly seen 
in $L$.  At 5 GHz, a central component that appears to be the core becomes 
prominent \citep{KKWJ98}, contrary to what is expected in the \citet{R83} model. 

\subsection{PSR B1822+00 = J1825+0004;~~~~P = 0\fs779;~~~~~~~~Fig.~14}

There are two components trailing our main component at 430 MHz:  one about
$6\degr$ later, and a faint one trailing the main component  by $10\degr$.  
Both trailing  components can be seen  at most
frequencies ranging from 610 to 1642 MHz [\citet{GL98}; \citet{W99}]. 
 The principal component has moderate $L$ with a flat position angle and 
strong $V$.  The fainter
trailing components do not have measurable polarization.  The 
classification is uncertain.

\subsection{PSR B1853+01 = J1856+0113;~~~~P = 0\fs267;~~~~~~~~Fig.~14}

This  pulsar is associated with the SNR W44. We detect linear polarization 
with a rather constant position angle under the primary component in our 430 
MHz profile. The 410 MHz profile of \citet{GL98} appears to be smeared, as it
is twice as wide as ours. Our 430 MHz polarized profile is rather similar 
to the 606 MHz profile of \citet{GL98}.  We detect a weak, unpolarized trailing 
component at $7\degr$ longitude. Profiles up to 1642 MHz reveal a
single form in total power, with unmeasurable polarization [\citet{GL98}; 
\citet{W99}].  The classification is uncertain.

\subsection{PSR B1854+00 = J1857+0057;~~~~P = 0\fs357;~~~~~~~~Fig.~14}

We observe moderate circular and linear polarization, edge depolarization, 
and a systematically rotating position angle at 430 MHz.  These results 
are similar to what was found  by \citet{W99} at 1418 MHz, who tentatively 
classified it as a conal
single $(S_d)$ on the basis of these properties.  The 430 MHz profile is
somewhat wider than the 1418 profile, as expected of  conal emission.  
Observations at even lower frequencies should reveal the bifurcation of the 
profile into the usual leading and trailing conal components. Note that
there is evidence in our profile for multiple components. Higher $S/N$
observations will help to specify the exact classification.  At present
the classification is unknown, although it should almost certainly include
conal emission of some type.

\subsection{PSR B1859+07 = J1901+0716;~~~~P = 0\fs644;~~~~~~~~Fig.~15}

The profile is an asymmetric single, with low--level leading and trailing 
shoulders of emission.  The main component retains essentially the same form 
in total power at frequencies up
to 1.6 and 4.85 GHz [\citet{GL98}; \citet{KKWJ98}].  Marginal detections of
linear polarization are reported at intermediate frequencies by \citet{GL98}.
Neither \citet{GL98} nor we detect significant polarization near 430 MHz.

\subsection{PSR B1906+09 = J1908+0916;~~~~P = 0\fs830;~~~~~~~~Fig.~15}

There are clearly two principal components with a separation of $\sim30\degr$.
Circular polarization changes sign between these two components, arguing
for a core in the saddle.  There is even the possibility of an additional 
faint component at $\sim50\degr$ longitude although it is only marginally
above the noise.  The 606 and 1408 MHz profiles of \citet{GL98} appear
to show only the leading component; presumably because the others are too weak
for them to detect.  We tentatively assign this to the Triple $(T)$ class.

\subsection{PSR B1907+12 = J1910+1231;~~~~P = 1\fs442;~~~~~~~~Fig.~15}

The profile shows an apparent scattering tail at 430 MHz, whereas it appears
symmetric at 1420 MHz \citep{seiradakis95}.  Some linear polarization and
a disordered position angle are apparent near the pulse centroid.

\subsection{PSR B1913+167 = J1915+1647;~~~P = 1\fs616;~~~~~~~~Fig.~15}

We observe a double form with moderate linear and circular polarization,edge 
depolarization, and a systematically sweeping position angle.   Other 
observations up to 1.6 GHz show the double form.  This is a double $(D)$ 
profile. The \citet{GL98} profile at
a similar frequency (410 MHz) shows the same form in total power but negligible
polarization.  Their polarized profile at 606 MHz actually looks more similar
to our 430 MHz result. The 430 MHz polarimetry of \citet{RB81} is quite
similar to ours except that they show a notch in $L$ just after the main
component.  It is hard to ascribe the difference to anything other than
temporal variations through the intervening decade. 

\subsection{PSR B1915+22 = J1917+2224;~~~~P = 0\fs426;~~~~~~~~Fig.~16}

Our 430 MHz data show a broad, unremarkable single profile with sweeping (low)
linear polarization near the central portions. We are unaware of profiles at 
other frequencies.

\subsection{PSR B1919+20 = J1921+2005;~~~~P = 0\fs761;~~~~~~~~Fig.~16}

This pulsar is a double $(D)$ in form.  Linear polarization is unmeasurably
weak.  There are no measurements at other frequencies.

\subsection{PSR B1920+20 = J1922+2018;~~~~P = 1\fs173;~~~~~~~~Fig.~16}

We observe three components in total power, each with significant $L$ but
depolarization at the profile outer edges.  The
classic S--shaped position angle sweep is evident.  The 410 and 606  MHz 
profiles of \citet{GL98} show the two main components in total power $(I)$
but their $S/N$ is insufficient to show significant polarized power.  Their 
1408 MHz data appear to show a narrowing of the overall profile to 
$\sim10\degr$ width.  This is a triple $(T)$.

\subsection{PSR B1921+17 = J1923+1705;~~~~P = 0\fs547;~~~~~~~~Fig.~16}

We have detected a three--component profile with low linear and sign--changing 
circular polarization under the central region.  In 
the absence of profiles at other frequencies, we tentatively classify this
as a Triple $(T)$.

\subsection{PSR B1922+20 = J1924+2040;~~~~P = 0\fs238;~~~~~~~~Fig.~17 }

Our 430 MHz profile is similar to \citet{GL98} at 610 MHz:  a broad single 
in form, with unmeasurable polarization.

\subsection{PSR B1925+18 = J1927+1855;~~~~P = 0\fs483;~~~~~~~~Fig.~17}

We detect two or three distinct components across $>30\degr$ of longitude.
Circular polarization changes sign under the central portion. Linear is
unmeasurably low.
The 610 MHz profile of \citet{GL98} appears fairly similar, given its lower
sensitivity.

\subsection{PSR B1926+18 = J1929+1846;~~~~P = 1\fs221;~~~~~~~~Fig.~17}

This pulsar was discovered to be a mode--changer by \citet{f81} in 430 MHz
Arecibo observations.  Our profile displays the normal mode.  The only new
information is a detection of moderate circular polarization in the first
component. The 21 cm profile of \citet{W99} was also too noisy to help
in classification, so we are left with the tentative Triple $(T)$ 
designation of \citet{R83}.

\subsection{PSR B1927+13 = J1930+1316;~~~~P = 0\fs760;~~~~~~~~Fig.~17}

\citet{W99} tentatively classified this pulsar as a triple $(T)$, based on 
their low--sensitivity 21 cm observations and early 
right circularly polarized observations at 430 MHz 
\citep{GR78}.  Our 430 MHz observations strengthen this classification, with
our clear detection of a leading component. The \citet{GL98} 610 MHz
profile may also show this leading component.  There is low linear 
polarization
throughout much of the profile except the leading and trailing edges, with
a fairly steady position angle, and some low level of circular 
polarization as well.

\subsection{PSR B1929+10 = J1932+1059;~~~~P = 0\fs226;~~~~~~~~Fig.~18}

This strong pulsar exhibits emission over most longitudes,  \footnote{In fact,
PSR B1929+10 is also a {\em{continuous}} radio source at 408 MHz \citep{PL85}.} 
so we display several regions of its profile  in separate panels.  Our 
results are similar to the Arecibo 430 MHz  observations of \citet{p90} and
\citet{rr97}, who both have higher $S/N.$

Multifrequency observations and classifications are discussed extensively in 
\citet{W99} and need not be repeated here.  On the basis of their rotating 
vector model $(RVM)$ polarization fits, \citet{EW01} conclude that its entire 
profile
results from a single, wide emission cone that is almost aligned with the 
spin axis.  In contrast, \citet{rr97} argue that the pulsar is an orthogonal
rotator with main and interpulse emission originating from opposite magnetic
poles.

\subsection{PSR B1929+15 = J1931+1536;~~~~P = 0\fs314;~~~~~~~~Fig.~18}

We tentatively classify this pulsar as  a triple $(T)$.  Linear and circular
polarization are negligible compared to the noise.

\subsection{PSR  B1939+17 = J1942+1747;~~~~P = 0\fs696;~~~~~~~~Fig.~19}

We find a  broad $(>40\degr$ longitude)  profile, with marginally measurable 
linear polarization.

\subsection{PSR B1942+17 = J1944+1757;~~~~P = 1\fs997;~~~~~~~~Fig.~19}

Our measurements show a double $(D)$ profile at 430 MHz. The principal
(trailing) component has significant $V$ and $L$, with a relatively constant
position angle.  The leading component also has large circular polarization.

\subsection{PSR B1944+22 = J1946+2244;~~~~P = 1\fs334;~~~~~~~~Fig.~19}

Our 430 MHz observations display two overlapping emission components in
total power $(I)$, each having significant $L$ and separated by an
orthogonal mode jump in position angle.  Circular polarization $V$ is
 moderate and negative throughout.  These measurements roughly reproduce those
of \citet{RB81} except that our $|V|$ is smaller.
The rather  noisy 1418 MHz profile of \citet{W99} is similar in total power
$I$ and linear $(L)$, but circular $(V)$ has become negligible at the higher
frequency.  Note that
both our 430 MHz profile and the 1418 MHz profile of \citet{W99} show 
marginal evidence for an additional emission component at $\sim-10\degr$
longitude. \citet{W99} classified this object as a core single $(S_{t})$ 
primarily on the basis of its 430 MHz polarized properties displayed in 
\citet{RB81}. However, with our current measurements of lower $|V|$ than 
theirs, a possibly systematically sweeping position angle, and especially
edge depolarization;  we find that it is more likely a conal single 
$(S_{d})$. 
 
\subsection{PSR B1951+32 = J1952+3252;~~~~P = 0\fs040;~~~~~~~~Fig.~19}

This young pulsar, which is associated with the supernova remnant CTB--80,
has been detected at radio, x--ray and $\gamma$--ray energies. 
Radio polarimetry has previously been presented by \citet{W99} at 1.4 GHz and 
by \citet{GL98} at several frequencies ranging from 0.4 to 1.4 GHz. 
Neither our strong linear polarization nor our corresponding weak negative
circular polarization at 430 MHz are seen in the \citet{GL98}
408 MHz profile.  Indeed their 610 MHz profile resembles our 430 MHz profile 
much more closely.  The \citet{GL98} and \citet{W99} 1.4 GHz profiles are 
similar to one another although the former group measured somewhat higher $L$.  
The main
trend, also remarked upon by \citet{W99}, is that the wide, boxy profile
at 1.4 GHz ($10\%$ width $\sim40\degr$) begins  to show an overlapping,
leading component that adds another $\sim20\degr$ of (unpolarized) emission
at 430 MHz.  

The 102 MHz total power profile of \citet{KL99} shows a similar
form, rather than bifurcating as expected if it were a ``classic'' conal 
single as  tentatively suggested by \citet{W99}.  The new work of \citet{MR02}, indicating that {\em{inner}} cones do not spread at low frequency, could 
explain this one discrepancy among evidence suggesting the conal single 
classification.

\subsection{PSR B2016+28 = J2018+2839;~~~~P = 0\fs558;~~~~~~~~Fig.~20}

See \citet{W99} for an extensive discussion of multifrequency
observations and morphological classification.  We concur that
this pulsar is a  conal single $(S_{d})$, as first proposed by \cite{R83}.  

\subsection{PSR B2020+28 = J2022+2854;~~~~P = 0\fs343;~~~~~~~~Fig.~20}

Our observations constitute the highest resolution 430 MHz measurements of
this pulsar.  Note the complicated and interesting polarization behavior
near the center of the trailing component, with orthogonal mode transitions
and large $V$.  Other observations are summarized in \citet{W99}. Those
authors support the Triple $(T)$ classification, as do we.

\subsection{PSR B2028+22 = J2030+2228;~~~~P = 0\fs631;~~~~~~~~Fig.~20}

We have much finer time resolution than the 430 MHz polarimetric observations  
of \cite{RB81}, but we are only able to measure a total
power $(I)$ profile.  As summarized by \citet{W99}, this pulsar 
appears to be a  triple $(T)$, although we find further hints of additional
components. The central peak is much stronger at 430 MHz than at 21 cm, as
expected of a core.

\subsection{PSR B2034+19 = J2037+1942;~~~~P = 2\fs075;~~~~~~~~Fig.~20}

The profile shows multiple components, including some very low--level
emission in the leading and trailing edges. At 21 cm, only our two
principal components were clearly visible \citep{W99}, although those
authors commented on a ``nascent'' third component on the trailing edge.
The sign--changing circular  polarization in the central portion at both
frequencies suggests core emission, while leading and trailing edge
depolarization suggests addition conal emission as well. There are
several orthogonal mode jumps in position angle in the principal
emission component. This is a Triple $(T)$ or Multiple $(M)$ pulsar.

\subsection{PSR B2044+15 = J2046+1540;~~~~P = 1\fs139;~~~~~~~~Fig.~21}

Our profile is similar to the 408 MHz result of \citet{GL98}, except that
the latter authors measure significant negative $V$ in the first component.
(Their 610 MHz profile more cloesly resembles ours.)  Circular polarization 
changes over to slightly positive at 21 cm \citep{W99,GL98}. The
position angle sweeps systematically across the profile; edge depolarization
is also seen. We concur with the earlier classification of Double $(D)$,
with evidence summarized in \citet{W99}.

\subsection{PSR B2053+21 = J2055+2209;~~~~P = 0\fs815;~~~~~~~~Fig.~21}

The very strong circular polarization in the trailing component detected
by \citet{W99} at 21 cm has essentially disappeared in our 430 MHz profile.
The multifrequency polarimetry of \citet{GL98} shows that $V$ is also 
significant at intermediate frequencies of 925 and 606 MHz.  Linear 
polarization is stronger at 430 
and 606 MHz than at the higher frequencies and edge depolarization
is evident in our profile.  The principal component separation flares
from $5\degr$ at 21 cm through 430 MHz, to $7\degr$ at 102 MHz \citep{KL99},
 as expected of conal emission.  We classify the pulsar as a Double $(D)$.

As noted by \citet{W99}, however, the strong circular polarization in
the trailing component at higher frequencies might suggest core emission 
as well.  However core emission should be most prominent at {\em{low}}
frequencies, in contrast to what is observed.

\subsection{PSR B2110+27 = J2113+2754;~~~~P = 1\fs203;~~~~~~~~Fig.~21}

We detect moderate linear and strong circular polarization at 430 MHz, 
as did \citet{GL98} at 408 MHz.  Note also the edge depolarization.
The circular remains rather strong at 
610 MHz but is fading by 925 MHz \citep{GL98}.  Interestingly, the 
position angle
sweeps in the opposite direction in  our 430 and 610 \citep{GL98} MHz 
observations than at higher frequencies of 925, 1400, and 1600 MHz 
(Gould \& Lyne 1998; Weisberg et al 1999).   In the context of the $RVM$ 
model, this behavior 
would require that the emission beam center move from one  to the other 
side of our line
 of sight as a function of observing frequency.  An alternative
explanation, supported by the  position angle's  irregular trajectory 
and
orthogonal mode jump at 430 MHz, is that the lower frequency sweep is
caused by a changing balance in the strengths of two emission modes, 
rather
than resulting from fundamental magnetospheric geometry.

We support the $(S_{d})$ classification; full evidence is presented
in \citet{W99}.

\subsection{PSR B2113+14 = J2116+1414;~~~~P = 0\fs440;~~~~~~~~Fig.~21}

Our 430 MHz profile is similar to the 21 cm profiles of \citet{RSW89} and
\citet{W99}, although we show the first clear evidence for an orthogonal mode
jump in the central, rather highly circularly polarized region.  The 
multifrequency observations of \citet{GL98} show similar results, given
their lower sensitivity.  While \citet{IMS89} show what appears to be a 
scattering
tail at 102 MHz, \citet{KL99} display at that frequency an enhancement of
the leading part of the profile, centered $\sim10\degr$ before the peak, and
{\em{no}} scattering tail.                                                                                                                                                                                                                        
This is a core single $(S_{t})$ pulsar, as stated by \citet{RSW89} and 
\citet{W99}.

\subsection{PSR B2122+13 = J2124+1407;~~~~P = 0\fs694;~~~~~~~~Fig.~22}

Our 430 profile, double in form with a central saddle, appears similar to
the 21 cm measurements of \citet{W99}. However, the linear polarization degree 
is higher at 430 MHz, and we detect moderate negative circular polarization 
throughout.  Edge depolarization is clearly present. This is a conal double $(D)$.

While the 430 MHz linear 
polarization measurements of \citet{BCW91} showed a position angle sweep  
opposite in direction to the 21 cm measurements of \citet{W99}, our 430 MHz 
measurements show the {\em{same}} sign of sweep as do the \citet{W99} 21 cm 
data. (See PSR B22210+29 below for the resolution of a similar discrepancy.)

\subsection{PSR B2210+29 = J2212+2933;~~~~P = 1\fs005;~~~~~~~~Fig.~22}

 The systematic position angle sweep and edge depolarization indicate the 
presence of conal emission. \citet{KL99} show that only two principal 
components, separated by $\sim20\degr$, are present at 102 MHz.  \citet{W99} 
noted that
the strong $L$ and $|V|$ of the final component at 21 cm  indicated that 
it might contain a superposed core. At 430 MHz, $|V|$ has shrunk in this
component although it still exhibits another trademark of core emission,
namely a change of sign of $V$.  \citet{GL98}
present a variety of  lower sensitivity profiles from 408 through 1642 
MHz, which also illustrate the trends discussed here.  We still classify 
this pulsar as a  Multiple $(M)$, while leaving open the question of the 
location of
the core.

As reported above for PSR B2122+13, the 430 MHz linear polarization 
measurements of \citet{BCW91} agree with ours except that  their position
angle sweep is apparently inverted.  Our sweep direction agrees with the
21 cm measurements of \citet{W99}. 

\subsection{PSR B2303+30 = J2305+3100;~~~~P = 1\fs576;~~~~~~~~Fig.~22}

\citet{W99} summarize the case for conal single $(S_{d})$ classification;
especially the observation of edge depolarization, and profile bifurcation at 
102.5 MHz \citep{IMS89}. The  only inconsistency mentioned by \citet{W99}
 was that the sign of circular polarization in the Arecibo 430 MHz 
measurements of \citet{RCB74} was opposite to that found at 21 cm.  Our 
current observations remove that problem by showing
that $V$ is indeed negative at 430 MHz, as at 21 cm. The observations of 
\citet{GL98} also show little variation in polarization properties at
several frequencies between 410 and 1408 MHz.

\section{CONCLUSIONS}

We have measured full--Stokes parameter profiles of 48 pulsars at 430 MHz from
Arecibo Observatory.  We used our polarized data and previously published
profiles to classify these pulsars according to the \citet{R83} morphological 
scheme. Most of our profiles represent the highest $S/N$ measurements 
available. 

Faraday Rotation Measures  determined for most of these pulsars were used to 
study the  spatial and temporal variations of the galactic
magnetic field.   We used the NE2001 electron density model \citep{CL02} to provide  
new  $DM$ based distances for all pulsars having measured $RM$s.  The resulting 
significant changes in the estimated locations of many pulsars, 
along with our new $RM$ measurements, enabled us to have new
insights into the galactic magnetic field structure in selected directions.
We mapped the nearby magnetic field reversal interior to the Sun as a
virtual null in $RM$ extending for $\sim7$ kpc in a region 0.5 kpc in width.
Other large--scale magnetic field structures were delineated. It was found that 
field maxima tend to occur {\em{along}} spiral arms in the regions we studied,
in contrast with earlier work that showed maxima {\em{between}} the arms.

Many $RM$s changed on a several--year timescale.  
Simultaneous Dispersion Measure $(DM)$ measurements showed that electron density 
variations were {\em{not}} responsible for these changes.  Rather, a
newly identified subtlety of pulsar quasi--orthogonal polarization mode emission
which leads to large apparent $RM$ variations in average--pulse measurements 
\citep{Ramach03}, is the most likely explanation.

\acknowledgements{We thank D. Bilitza for supplying the ionospheric and 
geomagnetic models IRI 95 and IGRF 95, T. H. Hankins and J. M. Rankin for 
providing contemporaneous dispersion measure data in advance of publication,
and Pei Zhuan for analysis assistance.  The suggestions of referee J. M. 
Rankin were quite helpful.  Our searches of the literature
were greatly aided by the European Pulsar Network on--line Data Archive and by
the SIMBAD database, operated at CDS, Strasbourg, France.
JMC was supported by NSF Grants AST 98--19931 and 02--06036 to
Cornell University; and KED, BK, JTG, and 
JMW were supported by by NSF Grants AST 95--30710 and 
AST 00--98540 to Carleton College.  Arecibo Observatory is operated by 
Cornell University under cooperative agreement with the NSF.} 

{}

\clearpage

\clearpage

\clearpage

\clearpage

\clearpage

\clearpage

\clearpage

\clearpage

\clearpage

\clearpage

\begin{figure*}
\caption{\scriptsize Polarized profiles for four pulsars at 430 MHz.  
For each pulsar, the upper panel displays total $(I)$, linearly 
polarized $(L)$, and  circularly polarized  
$(V = S_{\rm left} - S_{\rm right})$ flux densities, normalized to the
peak $I$.  Total flux density is always the highest curve. Total and 
linearly polarized flux density are solid lines. Circularly polarized
flux density is dashed, with each dash having a duration of one 
original sample, which corresponds to 1/1024 pulsar 
period or $0\fdg3516$ of longitude. The final instrumental resolution,
after boxcar smoothing (if any), is indicated by a horizontal bar labelled
``sample.''  Dispersion smearing across a single 156 kHz filter is
illustrated by a horizontal bar labelled ``DM.'' The r.m.s. noise level
in Stokes parameter $I$ (total power) is shown by a vertical bar labelled
``rms I."  The lower panel displays
the position angle of linear polarization. The position angle is 
plotted only at longitudes where $L \ge 2\sigma_{\rm L}$. In the case of
some particularly weak pulsars, $L, V,$ and/or the position angle are not
plotted.}
\epsscale{0.6}
\plotone{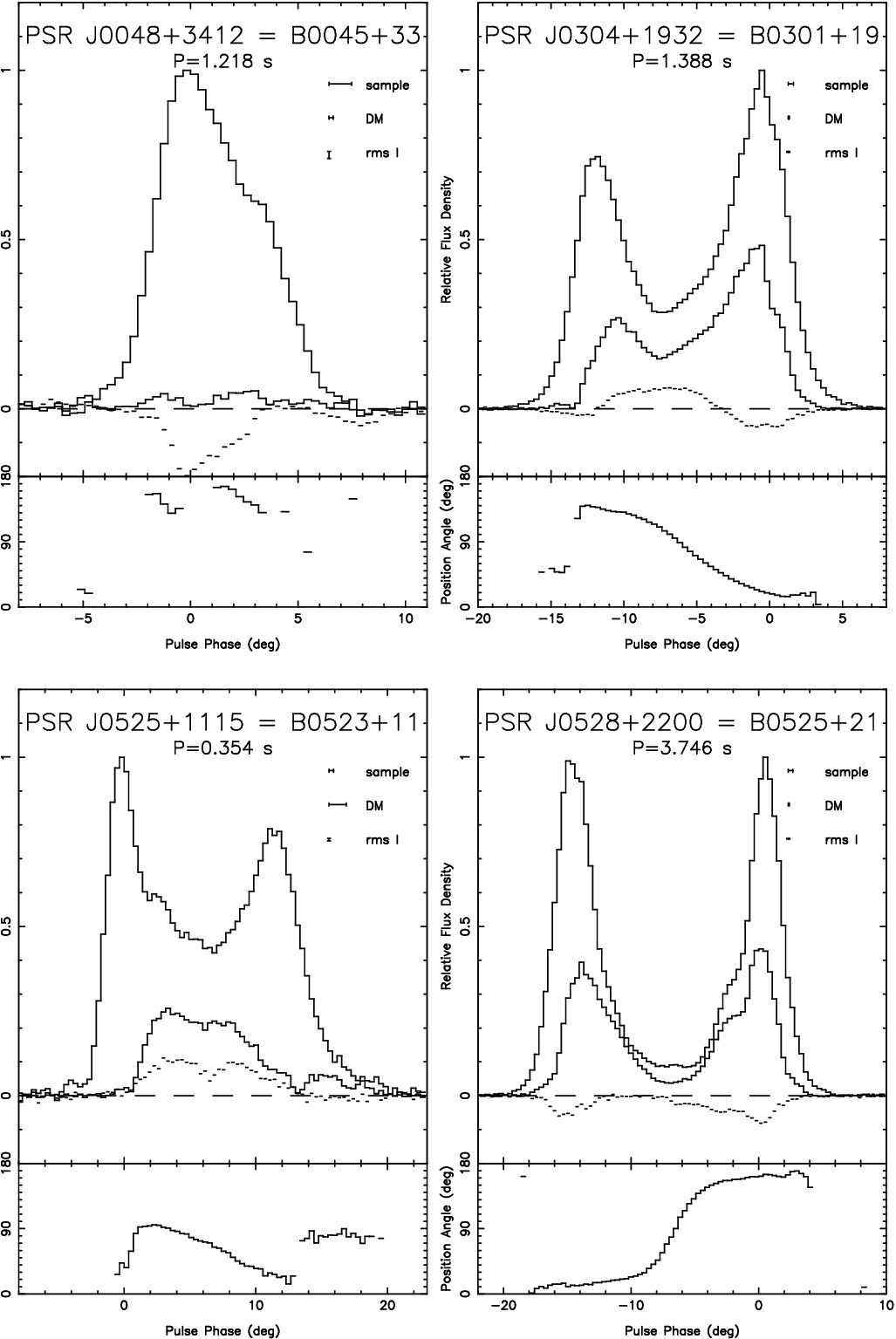}
\end{figure*}

\clearpage

\begin{figure*}
\caption{Polarized profiles for four pulsars at 430 MHz. 
See Fig. 10 caption for details.} 
\epsscale{0.7}
\plotone{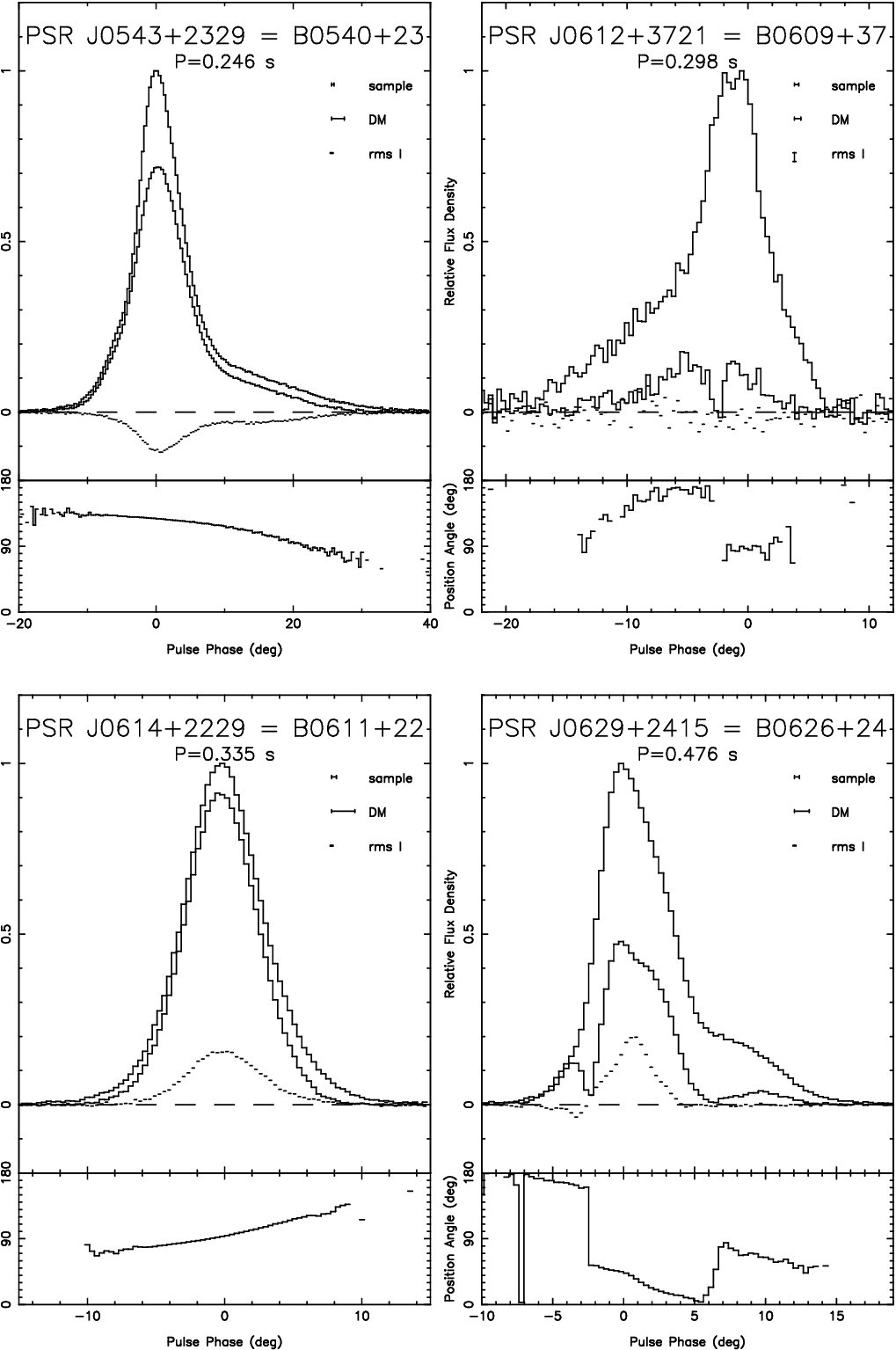}
\end{figure*}

\clearpage

\begin{figure*}
\caption{Polarized profiles for four pulsars at 430 MHz. 
See Fig. 10 caption for details.} 
\epsscale{0.7}
\plotone{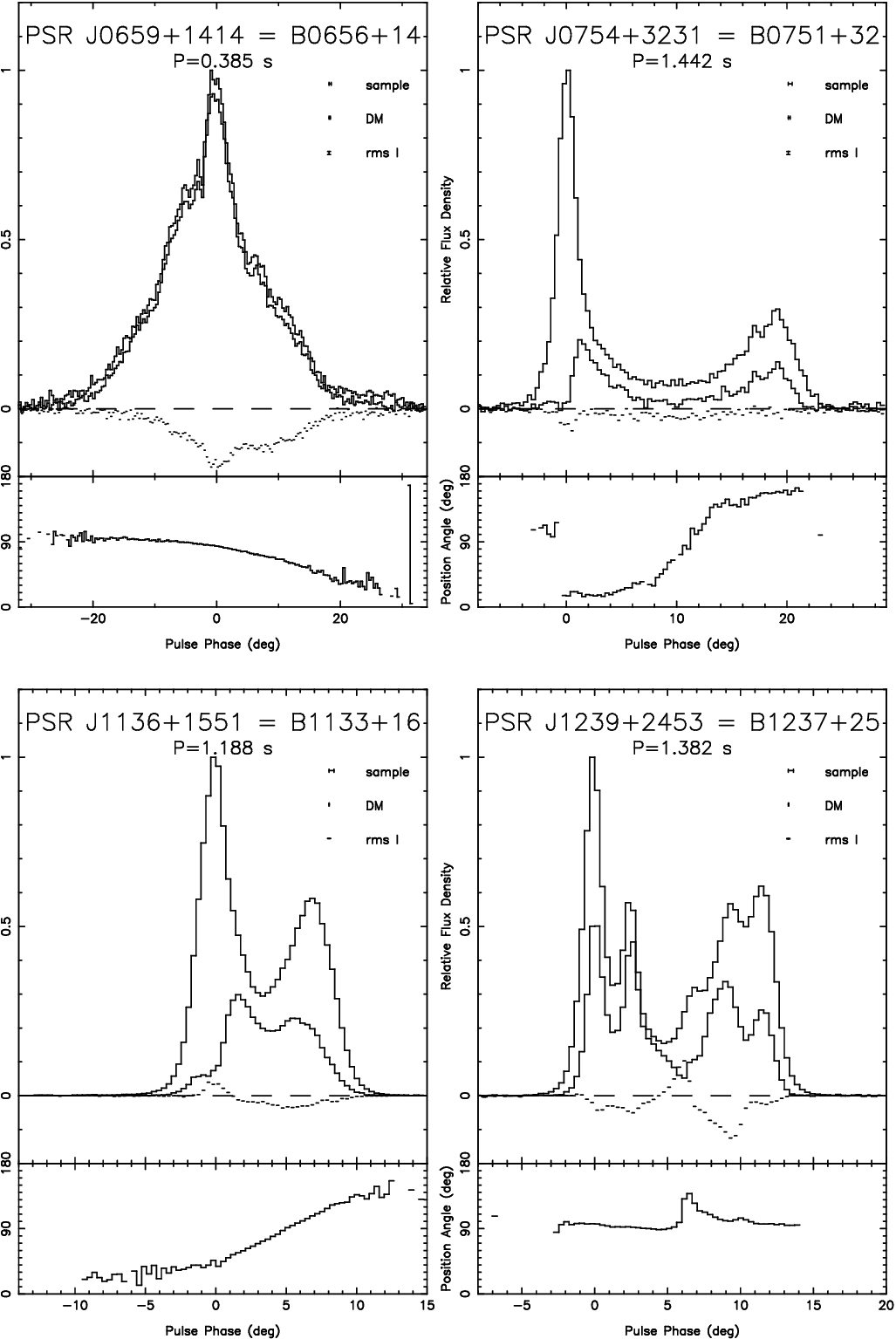}
\end{figure*}

\clearpage

\begin{figure*}
\caption{\scriptsize Polarized profiles for three pulsars at 430 MHz. The
main pulse and postcursor are displayed separately for PSR B1530+27. 
The position angle scale has the same origin in both panels on
this pulsar.   See Fig. 10 for additional details.}
\epsscale{0.6}
\plotone{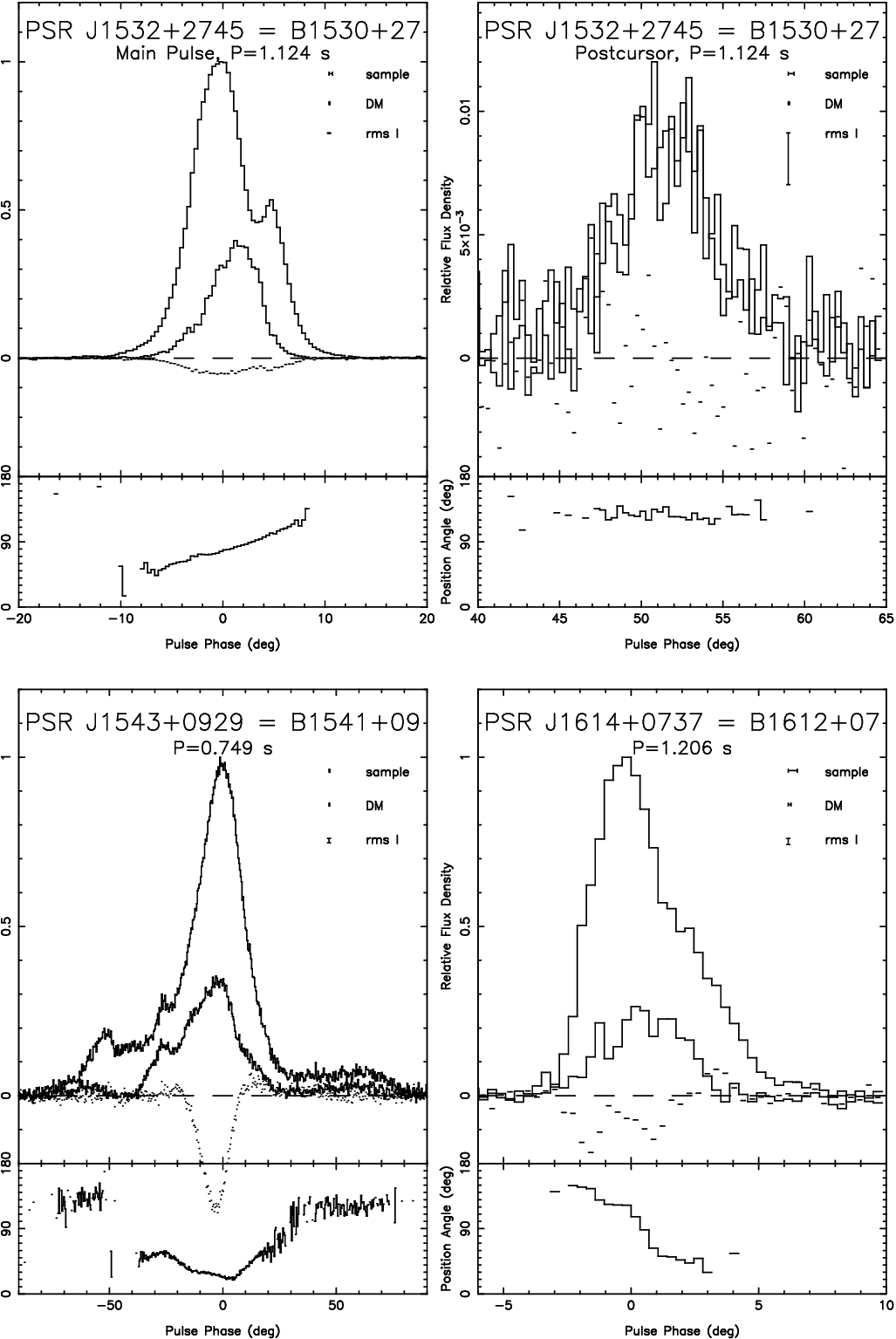}
\end{figure*}

\clearpage

\begin{figure*}
\caption{Polarized profiles for four pulsars at 430 MHz. 
See Fig. 10 caption for details.}
\epsscale{0.7}
\plotone{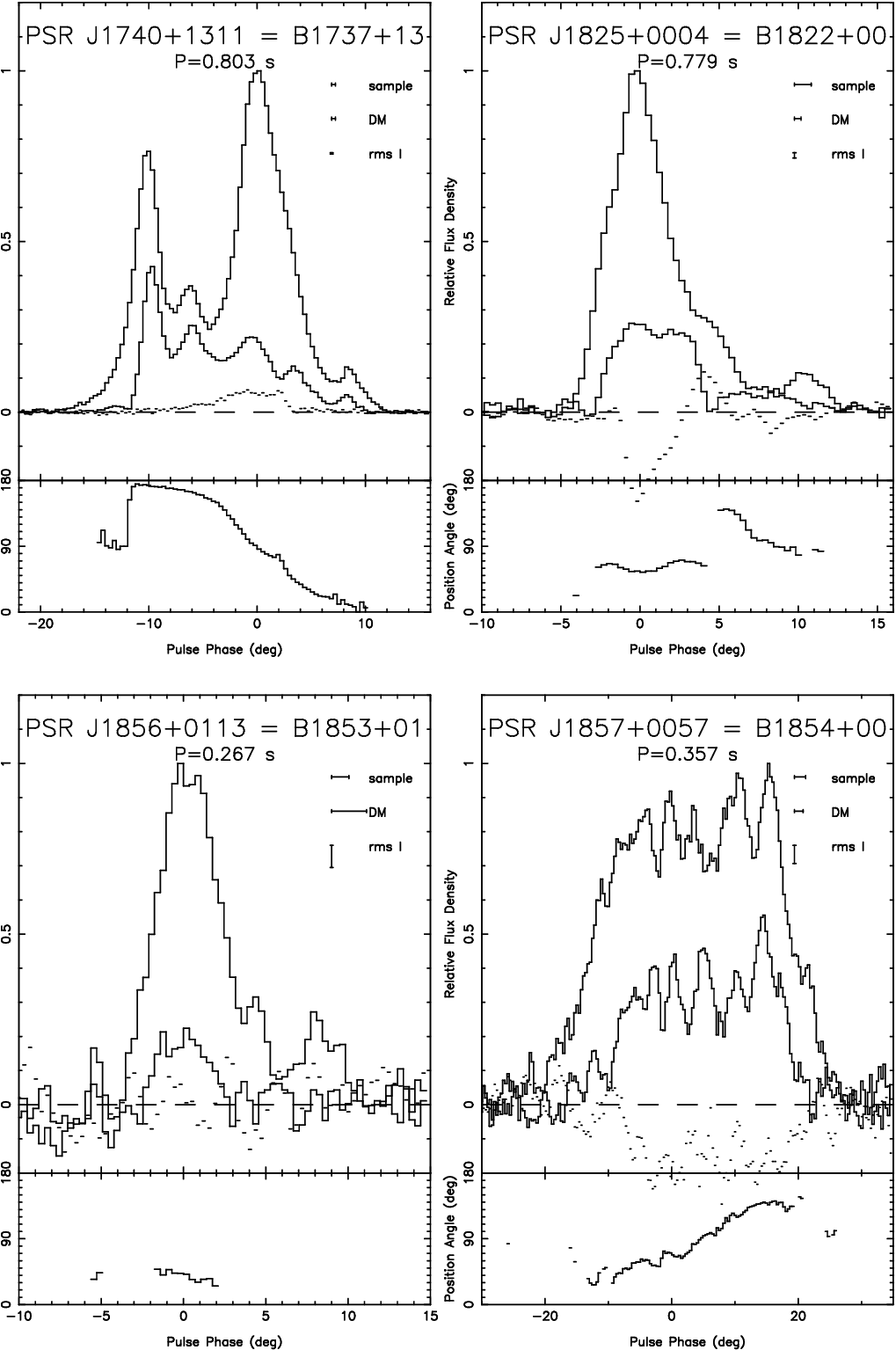}
\end{figure*}

\clearpage

\begin{figure*}
\caption{Polarized profiles for four pulsars at 430 MHz. 
See Fig. 10 caption for details.} 
\epsscale{0.7}
\plotone{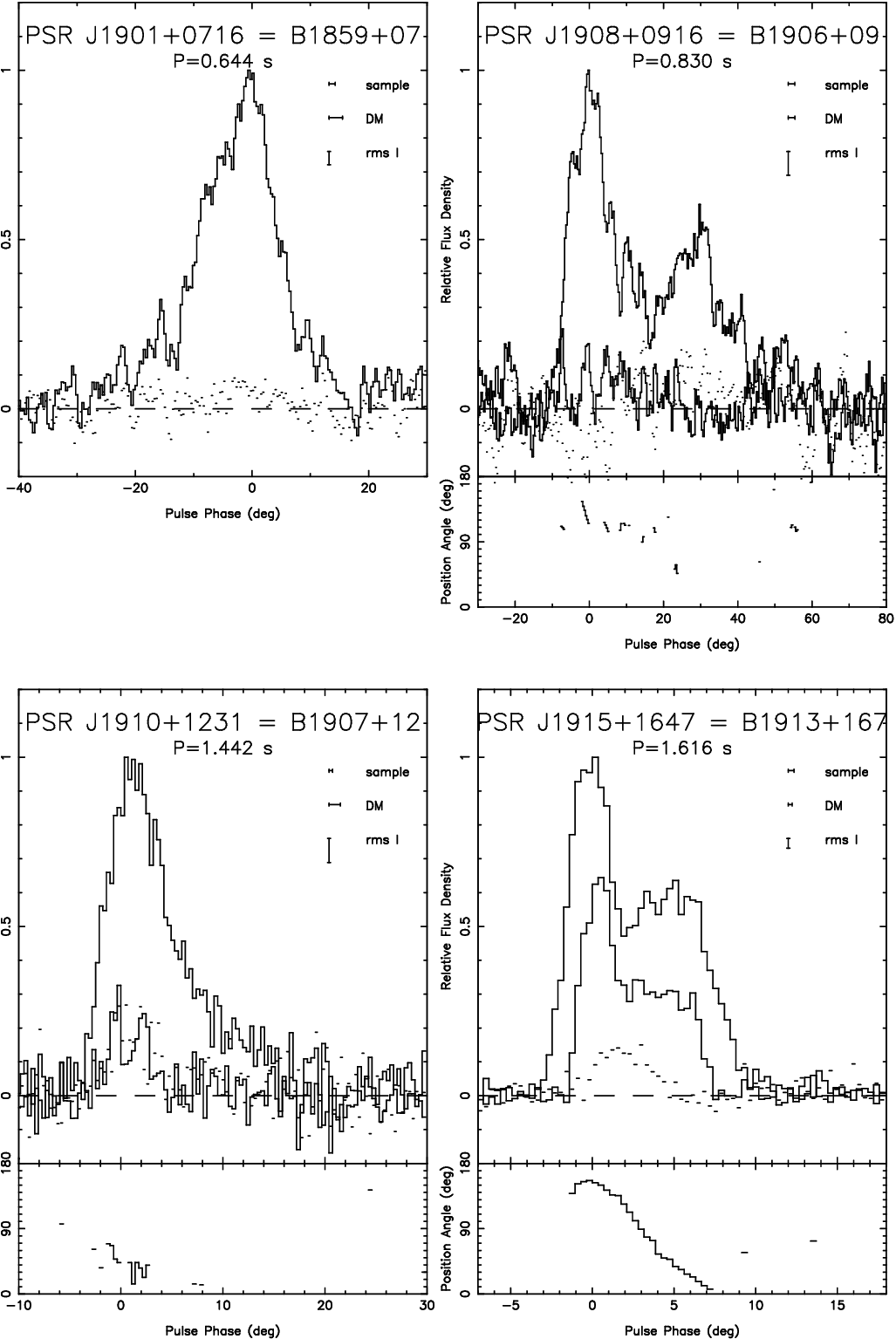}
\end{figure*}

\clearpage

\begin{figure*}
\caption{Polarized profiles for four pulsars at 430 MHz. 
See Fig. 10 caption for details.} 
\epsscale{0.7}
\plotone{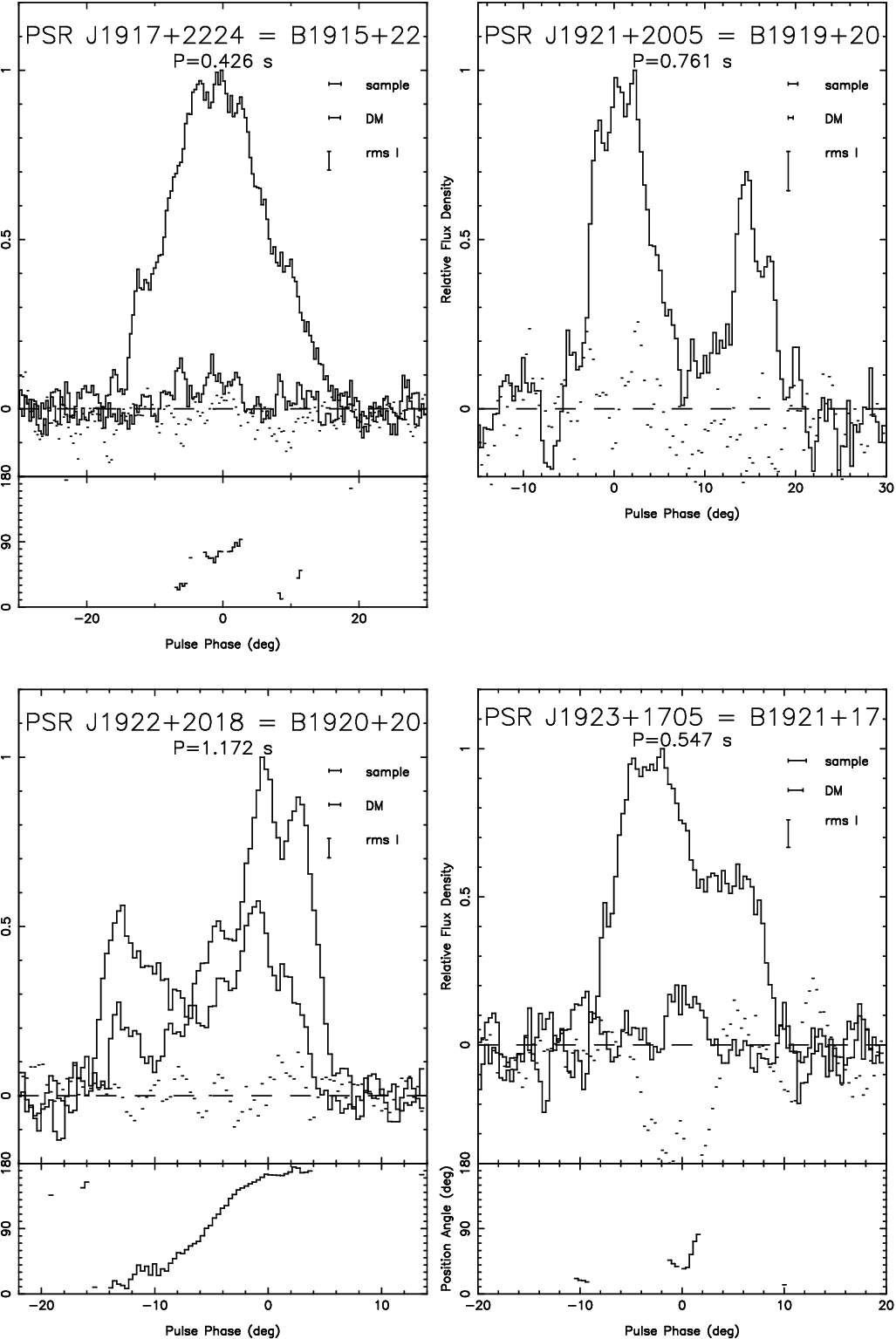}
\end{figure*}

\clearpage


\begin{figure*}
\caption{Polarized profiles for four pulsars at 430 MHz. 
See Fig. 10 for details.}
\epsscale{0.7}
\plotone{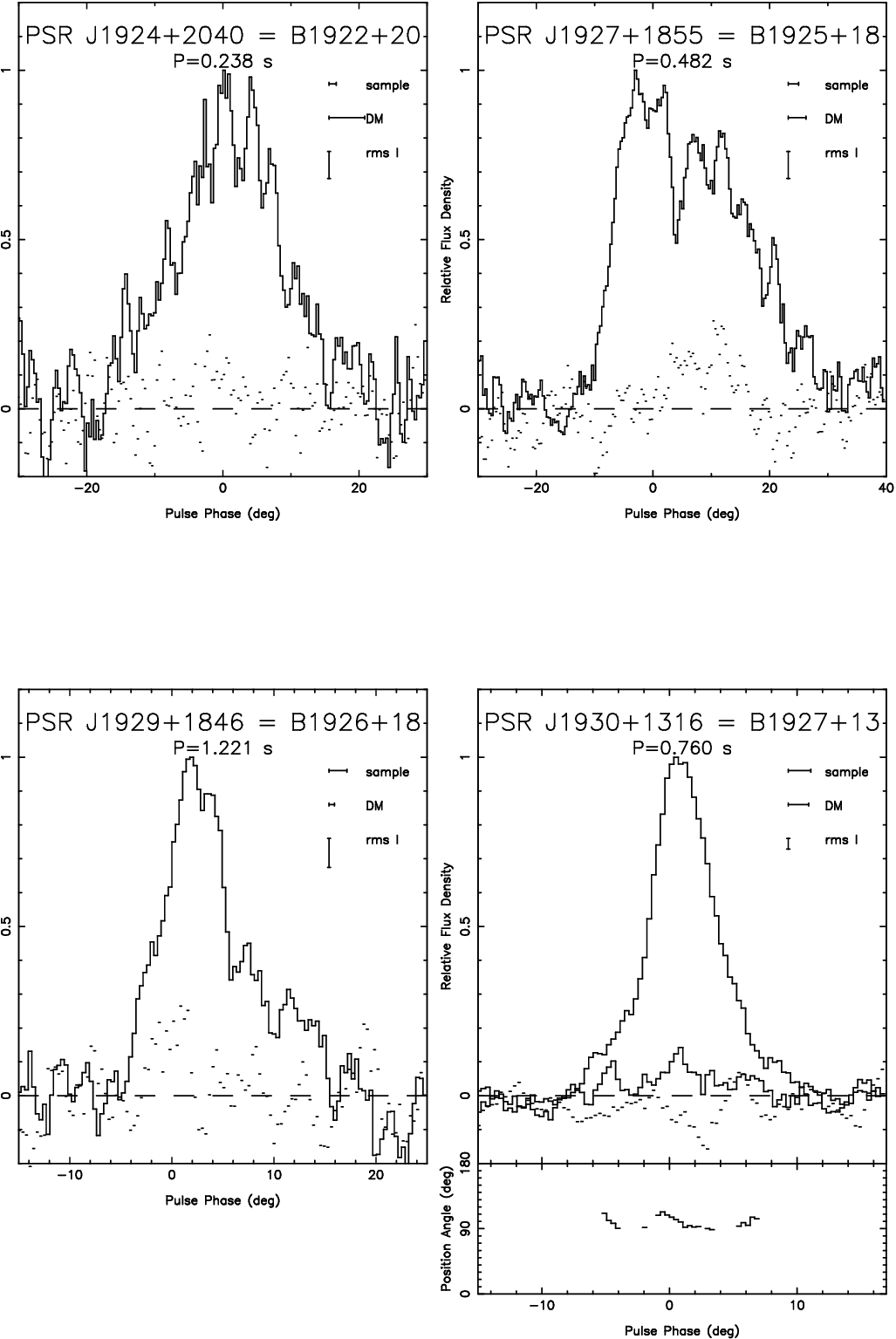}
\end{figure*}

\clearpage

\begin{figure*}
\caption{\scriptsize Polarized profiles for two pulsars at 430 MHz. 
The main pulse, postcursor, and interpulse  are displayed 
separately for PSR B1929+10. The position angle scale has the 
same origin in all three panels on this pulsar.   See Fig. 10 
for additional details.}
\epsscale{0.6}
\plotone{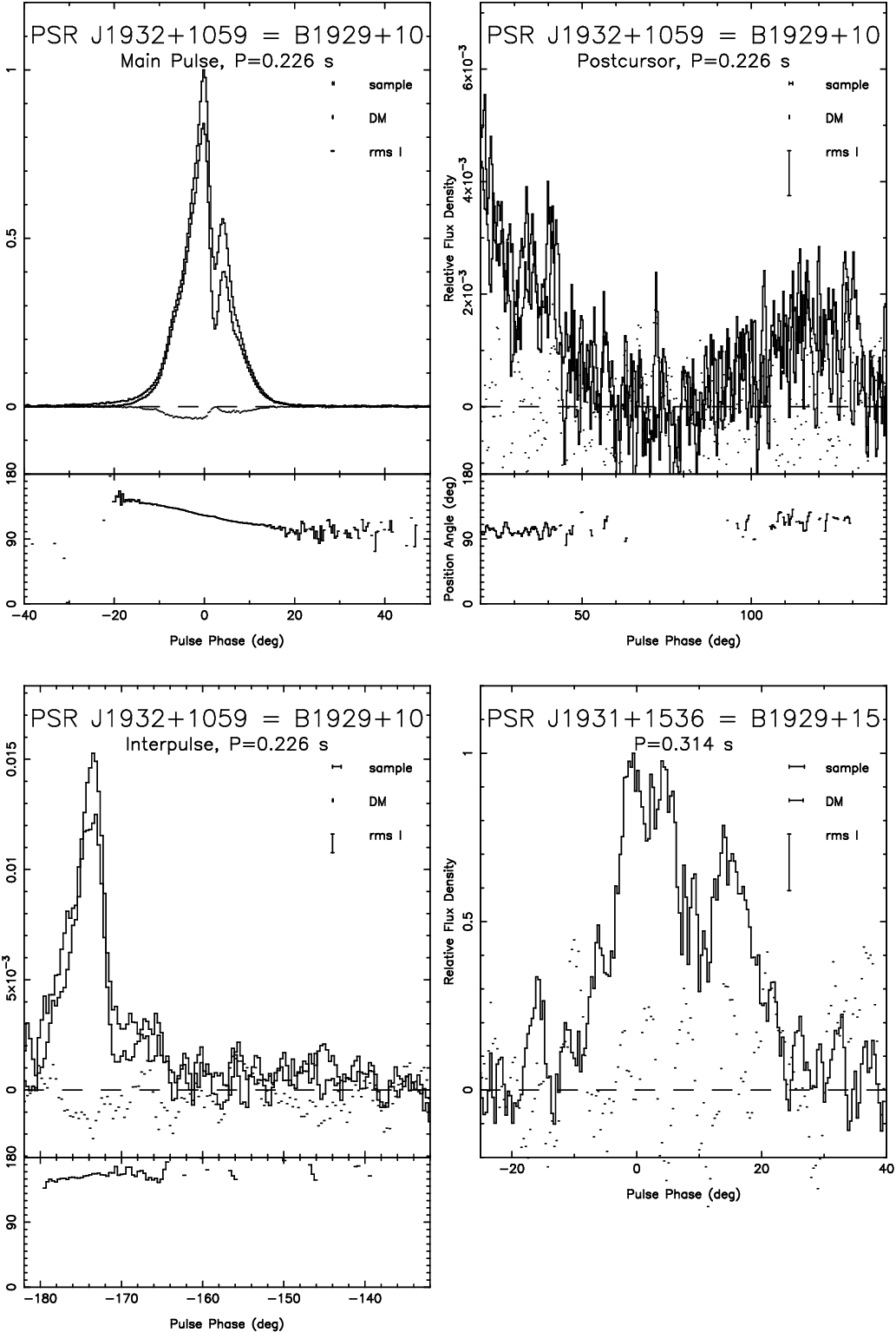}
\end{figure*}

\clearpage

\begin{figure*}
\caption{Polarized profiles for four pulsars at 430 MHz. 
See Fig. 10 for details.}
\epsscale{0.7}
\plotone{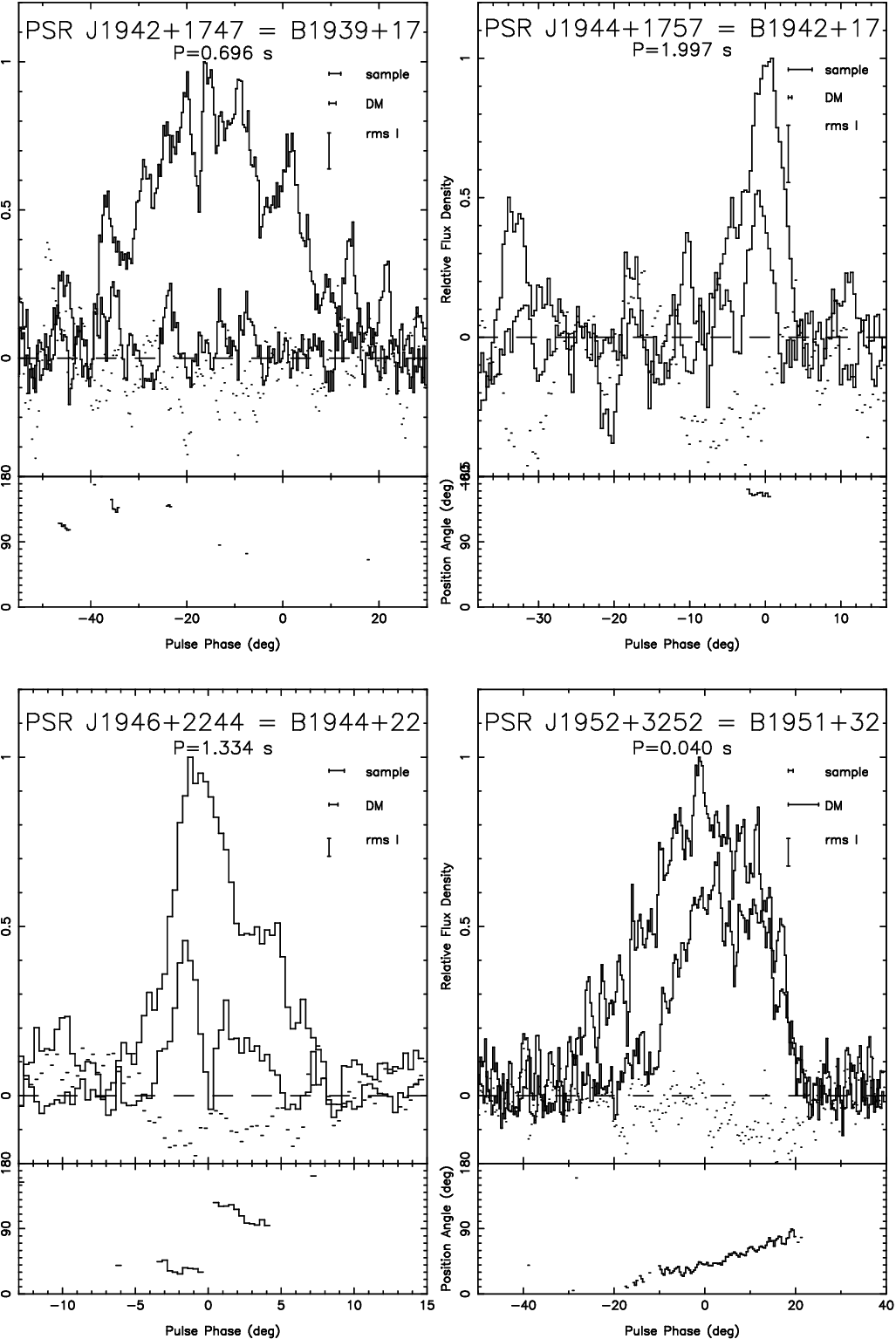}
\end{figure*}

\clearpage

\begin{figure*}
\caption{Polarized profiles for four pulsars at 430 MHz. 
See Fig. 10 for details.}
\epsscale{0.7}
\plotone{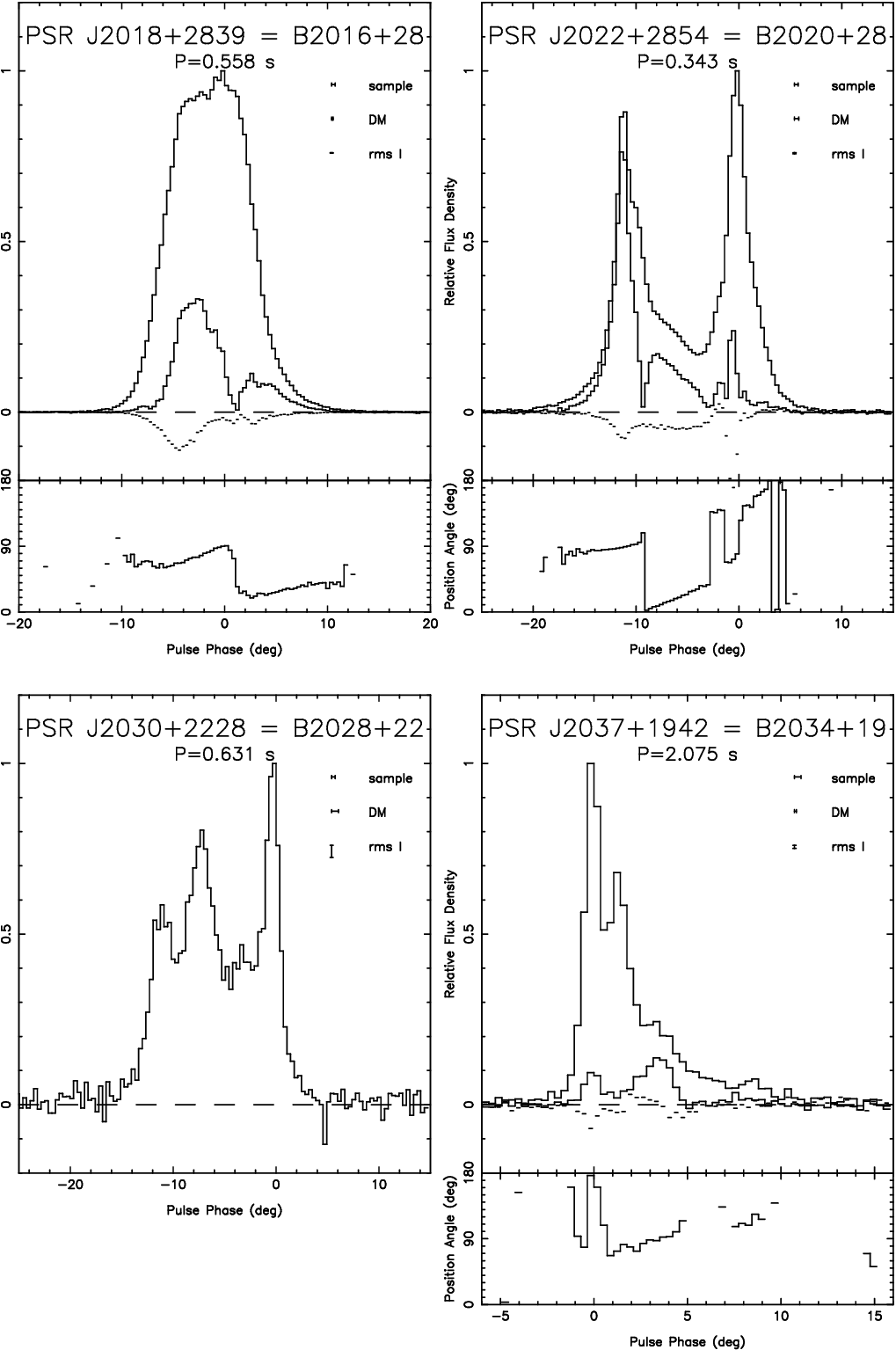}
\end{figure*}

\clearpage

\begin{figure*}
\caption{Polarized profiles for four pulsars at 430 MHz. 
See Fig. 10 for details.}
\epsscale{0.7}
\plotone{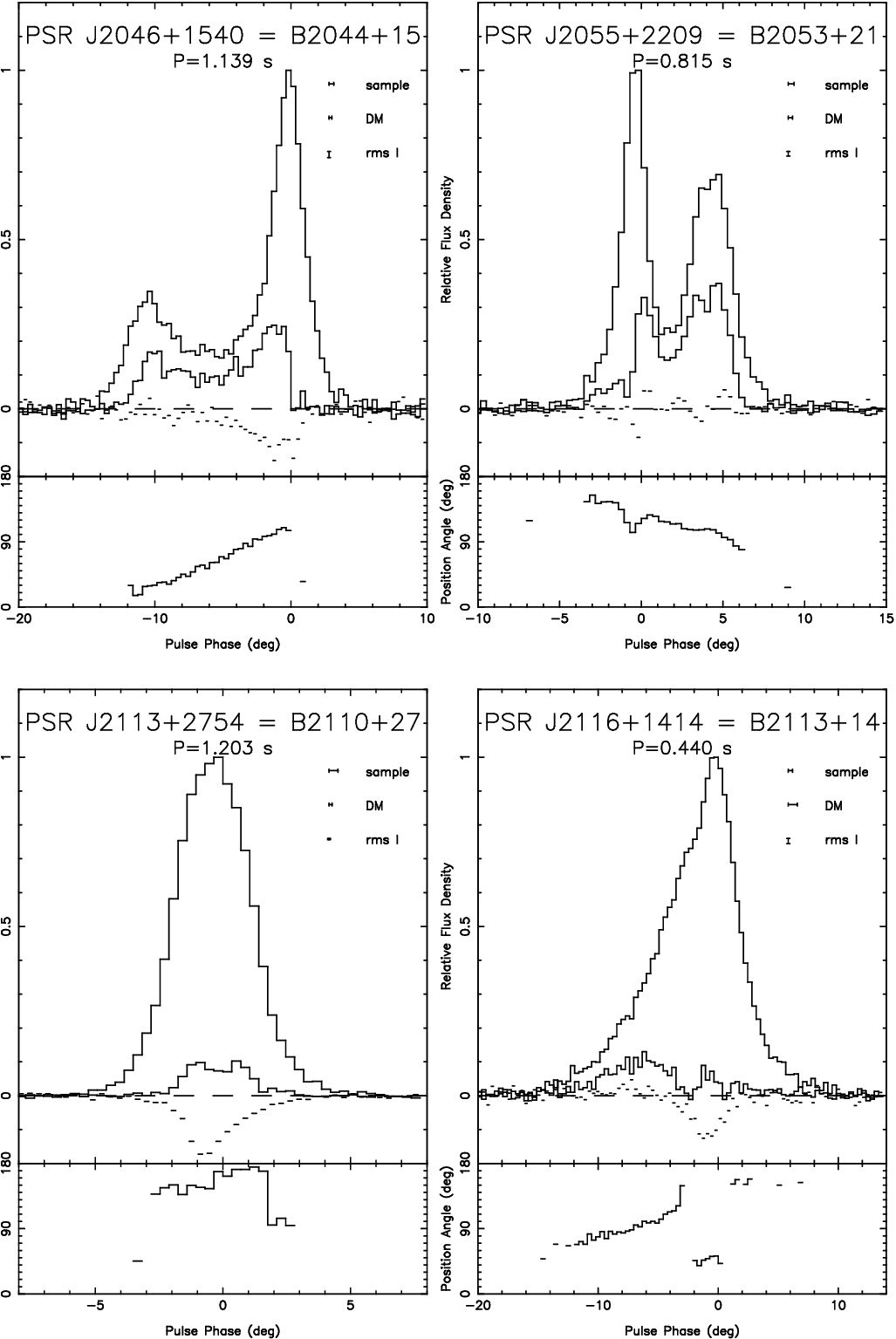}
\end{figure*}

\clearpage

\begin{figure*}
\caption{Polarized profiles for three pulsars at 430 MHz. 
See Fig. 10 for details.}
\epsscale{0.7}
\plotone{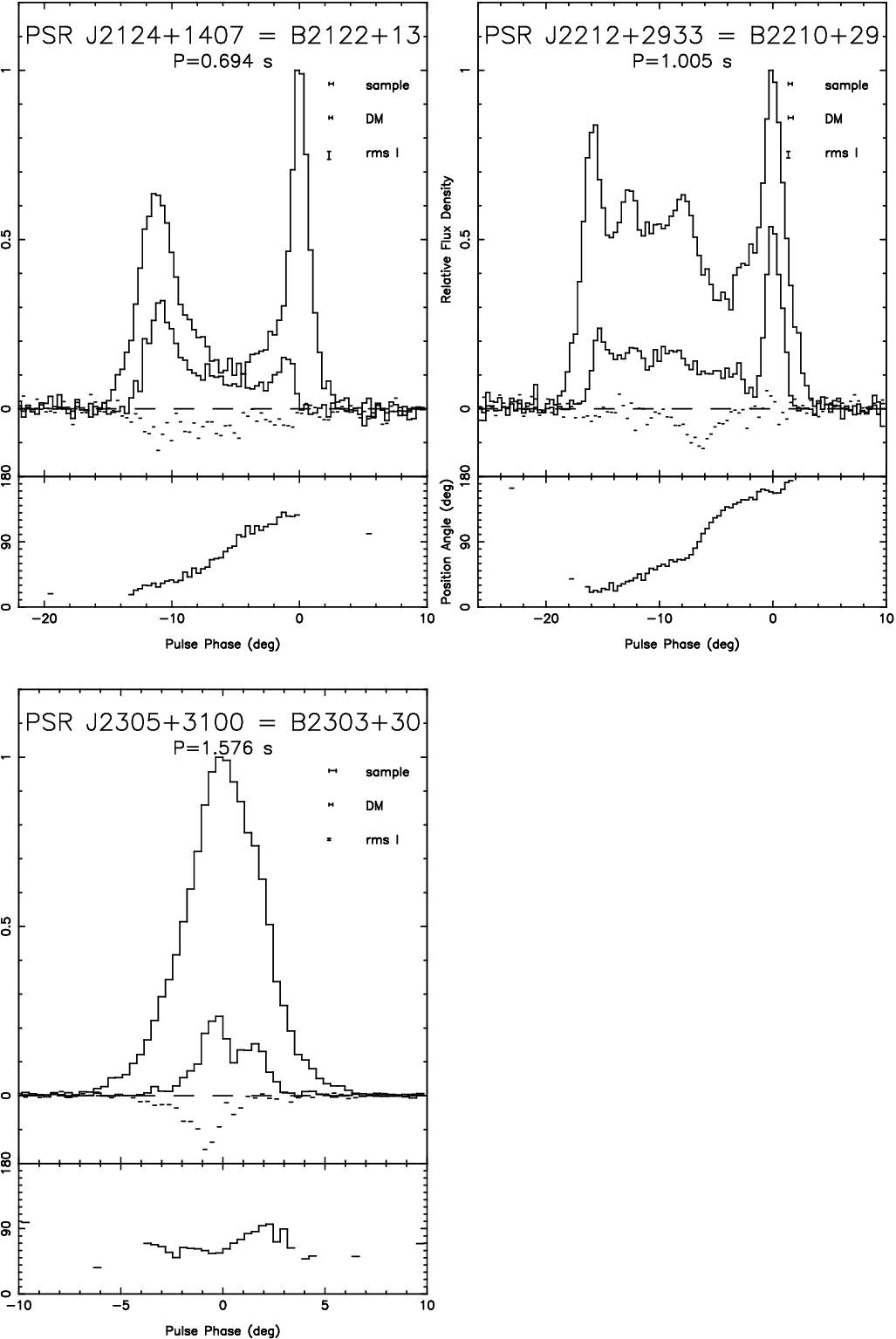}
\end{figure*}

\clearpage


\clearpage

\begin{table} 
\caption{ Rotation Measure Calibrators}
\scriptsize\begin{flushleft}
\begin{tabular}{llrlrllrl}
\tableline 
PSR J     &    PSR B &   Measured\tablenotemark{a} RM   &   
$\pm$\tablenotemark{b}  & Previous RM  & $\pm$  & Ref. & Difference &   $\pm$ \\
       &      & \multicolumn{2} {r}{(rad m$^{-2})$} & \multicolumn{2} {r}{(rad m$^{-2})$}  &  & \multicolumn{2} {r}{(rad m$^{-2})$} \\ 
\tableline
0528+2200   &    0525+21 &   -39.2   &      0.9 &  -39.6  &  0.2  &  \citet{m72}   &  0.4 & 0.9 \\
0614+2229   &    0611+22 &    67.0   &      0.3 &   67.0  &  0.7  &  \citet{hm81}  &  0.0 & 0.8 \\ 
1239+2453   &    1237+25 &    -0.4   &      0.6 &   -0.33 &  0.06 &   \citet{hm81} & -0.1 & 0.6 \\ 
1543+0929   &    1541+09 &    21.6   &      0.6 &   21.0  &  2.0  &   \citet{hl87} &  0.6 & 2.1 \\                      
1932+1059   &    1929+10 &    -5.9   &      0.6 &   -6.1  &  1.0  &   \citet{hm81} &  0.2 & 1.2 \\      
2022+2854   &    2020+28 &   -73.7   &      1.6 &  -74.7  &  0.3  &  \citet{m74}   &  1.0 & 1.6  \\                                              \tableline

\end{tabular}
\end{flushleft}
\normalsize
\tablenotetext{a}{The ``Measured'' $RM$ represents a closed--loop calibration 
test and should not be considered an independent measurement of $RM$.}
\tablenotetext{b}{The quoted uncertainty is twice the formal error.}
\end{table}

\newpage
\begin{table} 
\caption{Measured Rotation Measures}
\scriptsize\begin{flushleft}
\begin{tabular}{llrlrllrlr}
\tableline 
PSR J     &      PSR B &   Measured RM        &   $\pm$\tablenotemark{a}  & Previous RM  & $\pm$  & Reference & Difference & $\pm$ & $|$Difference$|$ \\
          &          & \multicolumn{2} {r}{(rad m$^{-2})$} & \multicolumn{2} {r}{(rad m$^{-2})$} & &  \multicolumn{2} {r}{(rad m$^{-2})$} & (\%) \\ 
\tableline

0304+1932  &   0301+19  & -5.7  &     1.1  &  -8.3  &    0.3  & \citet{m74}     &   2.6  &  1.1  &  31 \\
0525+1115  &   0523+11  &  35 &     3  &  29    &      5  &   \citet{hl87}  &   6  &  6 &  21 \\
0534+2200\tablenotemark{b}  &   0531+21  & -58 &     6  & -42.3 &    0.5  &   \citet{m72}    & -15  &  6 &  36  \\
0543+2329  &   0540+23  &   4.4 &     0.8  &   8.7 &    0.7  &  \citet{hm81}    &  -4.3  &  1.1 &  49 \\
0612+3721  &   0609+37  &  23 &     9                       \\                                                       
0629+2415  &   0626+24  &  69.5 &     2  & 82    &    4    &  \citet{hl87}    & -13  &  5 &  16 \\
0659+1414  &  0656+14   &  23.5 &     4  & 22    &    5    &  \citet{hl87}    &   1.5  &  6  & 7 \\
0754+3231  &  0751+32   &   4 &     6  & -7    &    5    & \citet{hl87}     &  11  &  8  & 157 \\
1136+1551  &  1133+16   &  8.9  &     0.8  &  3.9  &    0.2  &  \citet{m72}     &   5.0  &  0.8  & 128 \\
1532+2745  &  1530+27   &  1.0  &     3  & 54    &    11    &  \citet{hl87}   &  -53 & 11 & 98 \\
1537+1155  &  1534+12   & 10.6  &     2                        \\                                                       
1614+0737  &  1612+07   & 35    &     9    & 40     &      4   & \citet{hl87}   &  -5  &  10 & 13 \\
1635+2418  &  1633+24   & 31    &     4    & 31     &    5     & \citet{hl87}   &  0   &  7  & 0  \\
1740+1311  &  1737+13   & 64.4  &     1.6  & 73     &    5     & \citet{hl87}   & -9   &  5  & 12 \\
1825+0004  &  1822+00   & 21  &    13                       \\              
1854+1050  &  1852+10   & 502   &    25                        \\                                                       
1857+0057   &  1854+00  & 104 &    19                        \\                                                       
1901+0716   &  1859+07  & 321    &    38   & 282     &   13     &  \citet{RL94}  & 39   & 40 & 14 \\
1910+1231   &  1907+12  & 978 &    15                       \\                                                       
1915+1647   &  1913+167 & 172 &     3 & 161    &   11     &  \citet{hl87}  &11    & 11  & 7 \\
1917+2224   &  1915+22  & 192 &    49                        \\             
1922+2018   &  1920+20  & 301 &     7                          \\                     
1923+1705   &  1921+17  & 380   &   220                        \\                                          
1927+1855   &  1925+18  & 417   &    70                       \\                                         
1946+2244   &  1944+22  &   2 &    20                        \\                                          
1952+3252   &  1951+32  & -182 &    8                        \\                                          
2018+2839   &  2016+28  &  -27.3 &    2.1  & -34.6  &   1.4   & \citet{m72}    & 7.3   &  2.5  & 21 \\
2030+2228   &  2028+22  & -192 &    21                        \\                                          
2037+1942   &  2034+19  &  -97 &    10                        \\                                            
2046+1540   &  2044+15  & -100  &     5 & -101 &   6   &  \citet{hl87}  & 1   &  8  & 1 \\
2055+2209   &  2053+21  & -80.5  &     3                        \\                                          
2113+2754   &  2110+27  & -37  &     7 &  -65 &   8   & \citet{hl87}   & 28  & 11  & 43 \\
2116+1414   &  2113+14  & -26  &    11 &  -25 &   8   &  \citet{hl87}  &-1   & 13  & 4 \\
2124+1407   &  2122+13  & -57  &     8                         \\                                           
2212+2933   &  2210+29  & -168 &     5                         \\                                            
2305+3100   &  2303+30  &  -75.5 &     4 & -84  &   6   &  \citet{hl87}  & 9   & 7  & 11 \\

\tableline

\end{tabular}
\end{flushleft}
\tablenotetext{a}{The quoted uncertainty is twice the formal error.}
\tablenotetext{b}{The Crab Nebula Pulsar Rotation Measure includes a time--variable 
contribution from the Nebula \citep{RCIP88}.}
\normalsize
\end{table}

\newpage
\begin{table} 
\caption{Selected Dispersion Measures as a function of Time}
\scriptsize
\begin{flushleft}
\begin{tabular}{lllrrlrrr}
\tableline 
PSR J           &      PSR B &  Measured\tablenotemark{a} ~DM & Previous DM  & $\pm$                & Reference &Difference& $|$Difference$|$    &  \\
                &            &  {(pc  cm$^{-3})$}            & \multicolumn{2} {r}{(pc cm$^{-3})$} &            &(pc cm$^{-3})$& (\%)  &  \\ 
\tableline
J0304+1932	&	B0301+19	&	15.650	&	15.69	&	0.05	&	\citet{man1978}	&	-0.040	&	0.25	                 &  \\
J0528+2200	&	B0525+21	&	51.204	&	50.955	&	0.003	&	\citet{cra1970}	&	0.249	&	0.49	                 &  \\
J0614+2229	&	B0611+22	&	96.86	&	96.70	&	0.05	&	\citet{tay1975}	&	0.16	&	0.17	                 &  \\
J0629+2415	&	B0626+24	&	84.216	&	82.9	&	1.0	&	\citet{W81}	&	1.316	&	 1.6	                 &  \\
J1136+1551	&	B1133+16	&	4.8472	&	4.8479	&	0.0006	&	\citet{cra1970}	&	-0.0007	&	0.01	                 &  \\
J1239+2453	&	B1237+25	&	9.277	&	9.296	&	0.005	&	\citet{cra1970}	&	-0.019	&	0.20	                 &  \\
J1532+2745	&	B1530+27	&	14.67	&	13.6	&	1.0	&	\citet{W81}	&	1.07	&	 7.9	                 &  \\
J1635+2418	&	B1633+24	&	24.265	&	24	&	3	&	\citet{shi1980}	&	0.265	&	1.1	                 & \\
J1740+1311	&	B1737+13	&	48.73	&	48.4	&	1.0	&	\citet{ash1981}	&	0.33	&	0.68	                 &  \\
J2018+2839	&	B2016+28	&	14.1965	&	14.176	&	0.007	&	\citet{cra1970}	&	0.0205	&	0.14	                 &  \\
J2022+2854	&	B2020+28	&	24.623	&	24.62	&	0.02	&	\citet{mea1972}	&	0.003	&	0.01	                 &  \\
J2046+1540	&	B2044+15	&	39.71	&	38.8	&	1.0	&	\citet{W81}	&	0.91	&	2.35	                 &  \\
J2113+2754	&	B2110+27	&	25.122	&	24	&	2	&	\citet{shi1980}	&	1.122	&	4.7	& \\
J2116+1414	&	B2113+14	&	56.14	&	55.1	&	1.0	&	\citet{W81}	&	1.04	&	1.9	                 & \\
J2305+3100	&	B2303+30	&	49.575	&	49.9	&	0.2	&	\citet{m72}	&	-0.325	&	0.65	                 &  \\	
\tableline

\end{tabular}
\end{flushleft}
\tablenotetext{a}{From \citet{HR03}, for epoch 1988--1992, similar to our $RM$ measurement epoch.}
\normalsize
\end{table}

\end{document}